\RequirePackage{fix-cm}
\documentclass{svjour3}                     
\smartqed  
\journalname{EMSE}
\makeatletter \let\cl@chapter\relax \makeatother
\usepackage{cite}
\usepackage{amsmath,amssymb,amsfonts}
\usepackage{algorithmic}
\usepackage{graphicx}
\usepackage{textcomp}
\usepackage{balance}
\usepackage{xcolor}
\usepackage{comment}
\usepackage{adjustbox}
\usepackage{soul}
\usepackage[bookmarksopen=true]{hyperref}
\newcommand{\datadoiBOT}{%
  \begingroup\normalfont
  \smash{\href{https://doi.org/10.5281/zenodo.3610205}{\includegraphics[height=1.3\fontcharht\font`\B,trim=0 3 0 0]{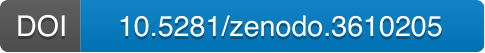}}}%
  \endgroup
}
\usepackage{booktabs}
\def\BibTeX{{\rm B\kern-.05em{\sc i\kern-.025em b}\kern-.08em
    T\kern-.1667em\lower.7ex\hbox{E}\kern-.125emX}}
\usepackage{cleveref}
\usepackage[title]{appendix}

\usepackage{listings}
\usepackage{color}

\definecolor{dkgreen}{rgb}{0,0.6,0}
\definecolor{gray}{rgb}{0.5,0.5,0.5}
\definecolor{mauve}{rgb}{0.58,0,0.82}

\begin{document}

\title{World of Code: Enabling a Research Workflow for Mining and Analyzing the Universe of Open Source VCS data}
\titlerunning{World of Code}
\author{Yuxing Ma \and Tapajit Dey \and Chris Bogart \and Sadika Amreen \and Marat Valiev \and Adam Tutko  \and David Kennard \and Russell Zaretzki \and Audris Mockus}

\institute{Yuxing Ma, Tapajit Dey, Sadika Amreen, Adam Tutko, David Kennard, Audris Mockus \at
              Department of Electrical Engineering and Computer Science,\\
              University of Tennessee, Knoxville, US \\
              \email{yma28@vols.utk.edu, tdey2@vols.utk.edu, samreen@vols.utk.edu, atutko@vols.utk.edu, dkennard@vols.utk.edu, audris@utk.edu}
          \and
          {Chris Bogart, Marat Valiev} \at
              Carnegie Mellon University, US\\
              \email{cbogart@andrew.cmu.edu, mvaliev@andrew.cmu.edu}
          \and
          Russell Zaretzki \at
              Department of Business Analytics \& Statistics,\\
              University of Tennessee, Knoxville, US \\
              \email{rzaretzk@utk.edu}
}

\date{Received: date / Accepted: date}

\maketitle

\begin{abstract}
  Open source software (OSS) is essential for modern society and, while
  substantial research has been done on individual (typically central)
  projects, only a limited understanding of the periphery of the
  entire OSS ecosystem exists. For example, how are the tens of millions of projects in the periphery interconnected through
  technical dependencies, code sharing, or knowledge flow?  To
  answer such questions we: a) create a very large and frequently updated collection of version control data in the entire FLOSS
  ecosystems named World of Code (WoC), that can completely cross-reference authors, projects, commits, blobs, dependencies, and history of the FLOSS ecosystems and b) provide capabilities to efficiently correct, augment, query, and analyze that data. Our current WoC implementation is capable of being updated on
  a monthly basis and contains over 18B Git objects. To evaluate its research potential and to create vignettes for its usage, we employ WoC in conducting several research tasks. In
  particular, we find that it is capable of supporting trend
  evaluation, ecosystem measurement, and the determination of
  package usage.  We expect WoC to
  spur investigation into global properties of OSS
  development leading to increased resiliency of the entire OSS ecosystem.
  Our infrastructure facilitates the discovery of key technical dependencies, code flow, and social networks that provide the basis to determine the structure and evolution of the relationships that drive FLOSS activities and innovation.
\end{abstract}

\section{Introduction}\label{intro}
Tens of millions of software projects hosted on GitHub and other
forges attest to the rapid growth and popularity of Free/Libre Open
Source Software (FLOSS). These online repositories include a variety
of software projects ranging from classroom assignments to
components, libraries, and frameworks used by millions of other
projects.  Such large collections of projects are currently archived in  public version
control systems, and, if made available for analysis, would represent a
unique opportunity to study FLOSS at large and answer both theoretical and practical questions that rely on the availability of the entirety of
FLOSS data. In particular, our infrastructure, referred to as World
of Code (WoC) and described below, supports a number of
  research and practical tasks that would not be possible without it.
For example, a census of open source software with types and
prevalence of projects, technologies, and practices and
could serve as a guide to setting policies or creating innovative
services. Also, the ability to discover complete chains of technical
dependencies, code flow, and global social networks of developers
to provide the basis to understand the structure and evolution of the relationships
that drive FLOSS activities and innovation. A sampling capabilities
of WoC provide a basis for
``natural experiments'' that evaluate the effectiveness of different
software development approaches.

WoC focuses on global reach and effective cross-referencing of public
version control data. The primary aim is to support research and industry
that studies/relies on open source ecosystems. Specifically,
it provides capabilities a) for stratified sampling in order to
improve external generalizability of empirical studies; b) to measure critical properties of code, dependencies, and
developer behaviour that are beyond reach of project-centric
approaches that can reach only data within a limited set of projects; c) to
study global networks of developers, code dependencies, and code
copying behaviours; d) for building tools that increase the transparency
and efficiency of open source in general and help mitigate risks,
thus allowing more use and contributions from industry.

Given the tremendous benefits from the collection of FLOSS
  development data, an infrastructure for mining FLOSS repositories
  and serving potential analysis and studies in software domain is
  in high demand. A number of other platforms have been built by
  researchers (see the Related Work section for details) to leverage
  this data for various ends. None of them, however, provide cross-referencing needed
  to measure code, developers, or dependencies in the context of the
  entire FLOSS.

We propose the following question:
\par \textbf{RQ: How to design an infrastructure
    to cross-reference source code change data over the entire FLOSS
    community in order to enable sampling, measurement, and analysis within and
    across software ecosystems?}

Our contribution is to describe a prototype of such an
infrastructure that can store the huge and growing amount of data in
the entire FLOSS ecosystem and can provide basic capabilities to
efficiently cross-reference, sample, and analyze it at that scale.
The primary focus is on the types of analyses that require global
reach across FLOSS projects. A good example is a software supply
chain~\cite{amreen2019methodology} where software developers correspond to the nodes or producers, relationships among software projects or packages represent
the ``chain'', and changes to the source code 
represent products or information (that flow along the chain) with  
corporate backers representing ``financing.'' It would be impossible to
measure properties and relationships of the producers without first having
data from a complete collection of software projects, since it is
impossible to know which projects each developer has contributed
to. Similarly, it is difficult to determine downstream dependencies, i.e., discovering all projects that depend on a specific project is not trivial. WoC is intended to
make such measurements straightforward.

Several formidable obstacles obstruct progress towards this vision.
The traditional approaches for obtaining the repository of a project
or a small ecosystem does not scale well and may require too many
resources and too much effort for individual researchers or smaller
research groups. Thus, the community needs a way to scale and share
the data and analytic capabilities. The underlying data are also
lacking in context necessary for meaningful analysis and are often
incorrect or missing critical attributes~\cite{M14}. Keeping such
large datasets up-to-date poses another formidable challenge.

In a nutshell, our approach is a software analysis pipeline starting
from discovery and retrieval of data, storage and updates, and
transformations and data augmentation necessary for analytic tasks
downstream. Our engineering principles are focused on using the
simplest possible techniques and components for each specific task
ranging from project discovery to fitting large-scale
models. The result is a prototype that appears to approximate the entirety of the publicly available
source code in version control systems and the latency of updates
on the existing hardware platform does not exceed one calendar month, which is relatively fast given the size of the dataset and the complexity of the task (See~\Cref{s:discovery,s:retrieval} for more details). Furthermore, we built a tool on top of the infrastructure and provided two types of API to enable wide data access for users.  

We begin with an overview of related work in
Section~\ref{s:related}. The architecture of the prototype
implementation of the infrastructure is discussed in Section~\ref{s:architecture}. 
We facilitate wide access to the large data collection by developing a tool on top of our infrastructure, which is described in Section~\ref{s:tool}, along with an evaluation of query performance.
We present a couple of applications in Section~\ref{s:applications}, demonstrating the tremendous value of this infrastructure to numerous software analytic tasks.
We also provide a tutorial about how to use the WoC infrastructure, using an example on Java language trend analysis in Section~\ref{s:tutorial}. We present a comparison between WoC and platforms offering similar functionality, viz. GHTorrent, Software Heritage, BOA, GH Archive, in Section~\ref{s:platform}. Details of a Hackathon event organized around the WoC infrastructure and projects undertaken in the event are described in Section~\ref{s:hackathon}, which demonstrates the communities' interest in WoC and some possible applications of the WoC infrastructure.
We discuss various ways of improving the existing infrastructure in Section~\ref{s:future}, discuss a few existing limitations in Section~\ref{s:limitations}, and conclude our paper in Section~\ref{s:conclusion}.

\section{Related Work}\label{s:related}

While we are not aware of a complete census of FLOSS with an
analysis engine, several large-scale software mining efforts
exist and may be roughly subdivided into attempts at preservation,
data sharing for research purposes, and construction of decision
support tools.

As described above, the aims of WoC is not to replace or replicate
any of these efforts, but to provide the cross-referencing needed to
analyze global properties of the entire FLOSS. Some of the design
decisions, for example, the use of Git object IDs, are intended to
make linking to and leveraging the information in other
systems easier or to simplify the provisioning of cross-referencing
services to enhance the capabilities of the other collections.

Software development is a novel cultural activity that warrants
preservation as a cultural heritage. The software source code, the
only representation of software that contains human readable
knowledge, needs to be preserved to avoid permanent loss of
knowledge~\cite{di2017software}. Software
Heritage~\cite{di2017software} is a distributed system involved in
collecting and storing large amount of open source development data
from various open source platforms and package hosts. It currently
has software from GitHub, GitLab, Debian, PyPI, etc., and contains 73M projects, 1.7B commits, and 15.6B source files.  
This effort does not presently focus on enabling applications to
software analytics. The provided APIs allows for quick query of every
historical particle in a software project and meets the preservation need,
however, it does not grant the access to the full relationships (e.g.,
the set of projects containing a given commit) among these particles
across entire collection of software. Quick access to these
relationships is crucial in conducting software analytics such as
identification of dependencies among artifacts and authors as well
as  code spread in the open source community.  

One potential value of archiving software lies in the reuse of software artifacts.
For example, Nexus~\cite{Nexus} repository manager, allows developers
to share software artifacts in a standard way and provides support
for building and provisioning tools (e.g. Maven) to access necessary
components such as libraries, frameworks and containers.

Commercial efforts, such as BlackDuck or
FOSSID\footnote{blackducksoftware.com,fossid.com} have proprietary
collections they use to determine if their clients have included
open source software within their proprietary software code. It is
generally not clear how complete these collections are nor if 
the companies involved might consider opening them for research purposes. 

In addition to source code and binaries, large scale collection of other software development resources could be
integrated with the source code data. For example,
GHTorrent~\cite{gousios2012ghtorrent,Gousi13,gousios2014exploratory,gousios2014dataset,gousios2014lean}
attempts to record every event for each repository hosted on GitHub
and provides multiple approaches (SQL request and MongoDB data dump)
for data access.  The primary limitation is that the collected
metadata is specific to GitHub and it does not include the
underlying source code as well.  Therefore, obtaining dependencies encoded within the
source code cannot be accomplished. A similar platform, named GH Archive\footnote{\url{https://www.gharchive.org/}}, is also focused on the collection of GitHub events. It provides new events dump per hour since 2011, and cloud service to meet the SQL based BigQuery.  FLOSSmole~\cite{howison2006flossmole} collects open source metadata
from various forges as a base for academic research but only focuses on
software project metadata.

Another platform is Candoia~\cite{tiwari2016candoia,tiwari2017candoia,upadhyaya2017accelerating,upadhyaya2018accelerating} which
provides software development data collections abstraction for
building and sharing Mining Software Repository (MSR) applications. 
In particular, Candoia contains many tools for artifact extraction from different VCSs and bug databases and it also support projects written in different languages. On top of these artifacts, Candoia created its general data abstraction for researchers to implement ideas and build tools upon. This design increased portability and applicability for MSR tools by enabling application on software repositories across hosting platforms, VCSs and bug recording tools. The approach is focused on the design and benefits of creating a specialized software repository mining language. While it abstracts a number of repository acquisition tasks, it also makes it more difficult to handle operational data problems that tend to occur at much lower levels of abstraction and tend to be too idiosyncratic for generalized abstraction.  The main drawbacks of Candoia are that it only supports limited programming language (JS and Java) based projects, and ecosystem-wide research might be difficult to implement since Candoia relies on users to provide software related data (e.g., targeted software repository URL) and eco-system wide compliance is generally low.

Other platforms are aimed at
improving reproducibility by providing a repository of datasets for
researchers to share their data.  These include PROMISE
Repository~\cite{Promise} and SourcererDB~\cite{ossher2009sourcererdb}.
PROMISE Repository is a collection of donated software engineering data. It was created to facilitate generations of repeatable
and verifiable results as well as to provide an opportunity for
researchers to extend their ideas to a variety of software
systems. Black Duck OpenHub is a platform that discovers open source
projects, tracks the development and provides the functionality of
comparison between softwares. Currently, it is tracking 1.1M
repositories, connecting 4.2M developers and indexing 0.4M
projects. SourcererDB is an aggregated repository of 3K open source
Java projects that are statically analyzed and cross-linked through
code sharing and dependency. On top of providing datasets, it also
provides a framework for users to create custom datasets using their
projects.  

Apart from providing datasets (repository) for potential users, platforms such as
Moose~\cite{ducasse2005moose},
RepoGrams~\cite{rozenberg2016comparing},
Kenyon~\cite{bevan2005facilitating},
Sourcerer~\cite{bajracharya2014sourcerer}, and Alitheia
Core~\cite{gousios2009alitheia} are more focused on facilitating
building and sharing MSR tools. Moose is a platform that eases
reusing and combining data mining tools. RepoGrams is a tool for
comparing and contrasting of source code repositories over a set of
software metrics and assists researchers in filtering candidate
software projects. Kenyon is a data platform for
software evolution tools. It is restricted 
to supporting only software evolution analysis. Sourcerer is an
infrastructure for large scale collection of open source code where
both meta data and source code are stored in a relational database. It
provides data through SQL query to researchers and tool builders
but is only focused on Java projects. Alitheia Core is a platform with
a highly extensible framework and various plug-ins for analyzing
software on a large database of open source projects' source code,
bug records, and mailing lists.  

Furthermore, there were efforts to
standardize software mining data description for enhanced
reproducibility~\cite{TaRe}.  None of the listed platforms focus on
both collection and analysis of the dependencies of the entirety of
FLOSS source code version control data.  Further, they contain either limited
collections (e.g. only GitHub, no source code, have only donated
data, or do not contain an analysis engine). For example, it is not
possible to answer simple questions such as ``In which projects has a
file been used?'', ``What projects/codes depend on a specific module?'', ``What changes has a specific author  made?'' etc.

Some large companies have devoted substantial effort to develop
software analysis platforms for the entire enterprise, aiming 
to improve the quality of software they build and to help the enterprise
achieve its business goals by providing recommendations to
software development organizations/teams, monitoring software
development trends, and prioritizing research areas. For
example, Avaya, a telecommunications company, built a
platform~\cite{HMPQ10}, which collects software development related
data from most of its software development teams and third parties
and enabled systematic measurements and assessments of the state of
software. CodeMine~\cite{czerwonka2013codemine},
is a software platform developed by Microsoft that collects a variety
of source code related artifacts for each software repository inside
Microsoft. It is designed to support developer decisions and provide
data for empirical research. We hope that similar benefits can be
realized with the WoC platform targeted to the entire FLOSS community.

Large scale software mining efforts also include domain specific
languages. Robert Dyer et al. developed
Boa~\cite{Dyer-Nguyen-Rajan-Nguyen-13,Dyer-Rajan-Nguyen-13,Dyer-13,boa-website,Dyer-Nguyen-Rajan-Nguyen-15,Dyer-Nguyen-Rajan-Nguyen-asd},
both as a domain specific language and as an infrastructure, to ease
open source-related research over large scale software
repositories. The approach is focused on the design and
benefits of an infrastructure and language combination.
However, the lack of explicit tools to deal with operational data
problems make it of limited use to achieve our aims.
Their collection procedures -discovery, retrieval, storage,
update, and completeness issues (for example, only certain languages
are supported)- are not the primary focus of this effort. 
The tools to deal with operational data problems common in version
control data are also lacking in Boa.


The system described in this paper is loosely modeled after a system
described a decade ago~\cite{M09msr,M07}. In comparison, at that time, Git
was just beginning to emerge as a popular version control system, but
now it dominates the FLOSS project landscape. The number of software
forges and individually hosted projects was much larger then in
contrast to the consolidation of forges and the overwhelming
dominance of GitHub. Furthermore, the scale of the FLOSS ecosystem
is more than an order of magnitude larger now and it continues to experience 
very rapid growth. WoC could not, therefore, reproduce that design closely 
and, instead, is focused on preserving the original Git objects and
on creating a design that enables both efficient updating of this huge database and ways to
cross-reference it so that the complete network of relationships among code and
people is readily available.

\section{Building the WoC Infrastructure}\label{s:architecture}

The process of mining individual Git repositories is
complex to begin with ~\cite{bird2009promises}, but becomes even more difficult on
a large scale~\cite{gorton2016software}. Specifically, using
operational data from software repositories requires resolution to
three major problems~\cite{M14}: the lack of context, missing
attributes or observations, and incorrect data. 

The lack of context: Operational data originates from traces collected and integrated from a variety of operational support tools. Each event recovered from such traces has a specific context of what may have been on the actor’s conscious and subconscious mind, the purpose of that action, the tools and practices used, and the project or the ecosystem involved. Each event, therefore, may have a unique meaning. Some actions, such as verbal communication, may be missing if conducted without tool support. Mining software traces in VCSs has the limitation of losing information that was not recorded by VCS. We, therefore, focus on revealing as much context as possible by providing related development components to each key property/object in the software repository through easy query on the map architecture described in Sec. \ref{maps}

Missing attributes or observations, and incorrect data: Data filtering or tampering may be done by the actors, operating tools, or data processing and integration at the time of action or later. Determining how to properly segment, filter, augment, and model such data to ensure that they contain a representative sample of relevant activities, is the fundamental challenge of engineering operational data solutions and it will be essential to develop methods to draw valid conclusions from such disparate and low-veracity data. In WoC implementation, we strictly followed the usage of the related VCS(Git) APIs for the extraction of development data to make sure we do not miss any attributes or observations. A common data quality issue with VCS is developer name disambiguation, and we detailed the correction of developer ID errors in Sec. \ref{identitymatching}

Moreover, to cope with these big data challenges we employed both vertical and horizontal
prototyping~\cite{rosch1999principles,agrawal2004integrating,lichter1994prototyping,budde1992prototyping} before building the complete infrastructure.

\begin{figure*}[htbp]

\centerline{\includegraphics[height=3.5cm]{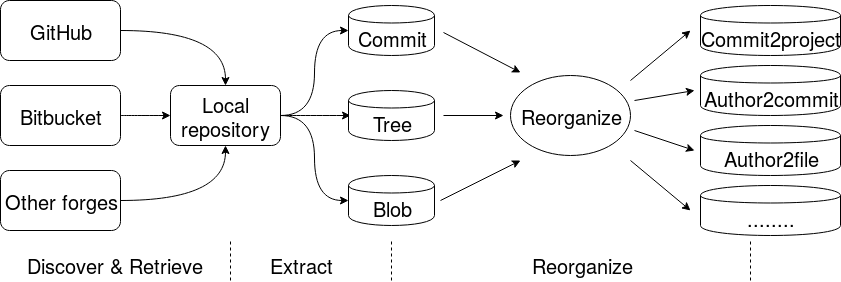}}

\caption{Overarching data flow}
\label{fig_flow}

\end{figure*}

In this section, we present a prototype WoC implementation.
It has three stages: project discovery, data retrieval, and
reorganization as shown in Figure~\ref{fig_flow}, which is typical of most big data systems, that use the layered data approach 
where the initial layers accumulate and process raw data and the
later layers produce cleaned/augmented data.  We also perform
  data  augmentation on the collected data, focusing on tasks like
  fork resolution~\cite{spinellis2020dataset} and author
  identity resolution~\cite{amreen2020alfaa,fry2020dataset}. 

The paper describes a rapidly evolving WoC prototype with some 
  aspects of the system evolving over time. While an attempt was made
  to focus on aspects that should persist, the need to provide
  specific examples almost ensures that the actual system will
  differ in some ways from the description provided here.  

\subsection{Project Discovery}\label{s:discovery}

Millions of projects are developed publicly on popular collaborative
platforms/forges such as GitHub, Bitbucket, GitLab, and SourceForge. Some of the
FLOSS projects can be identified from the registries maintained by
various package managers (e.g. CRAN, NPM) and Linux distributions
(e.g. Debian, Fedora). Most other project repositories, however, are
hosted in personal or project-specific sites. A complete list of
FLOSS repositories is, therefore, difficult to compile and maintain
since new projects and forges are constantly being created and many older forges
disappear continually. There is also a tendency for the FLOSS repositories to migrate to
(or be mirrored on) several very large forges~\cite{ma2016crowdsourcing}. A
number of older forges provide
convenient approaches to migrate repositories to other viable forges
before being shut down.
This consolidation has alleviated some of the challenge of discovering all
FLOSS projects~\cite{M09msr}, though the task remains nontrivial.
We discuss several approaches to project discovery below. To package
our project discovery procedure we have created a ``docker'' container\footnote{https://github.com/ssc-oscar/gather} that
has the necessary scripts.

\textit{Using Search API:} Some APIs may be used to 
 discover the complete collection of public code
repositories within a forge. The APIs are specific to each forge and 
come with different caveats. Most APIs tend to be rate limited (for
user or IP address) and the retrieval can be sped up by pooling the
IDs of multiple users. 

\textit{Using Search Engine:} Search engines (e.g. Google or Bing) 
can supplement the discovery of FLOSS project repositories on
collaborative forges when the forge does not provide an API, or when the API is broken. 
The primary drawback is the incompleteness
of the repositories discovered.

\textit{Keyword Search:} Some forges provide keyword based search of
public repositories, which is a complementary approach when a
forge does not provide APIs for the enumeration of repositories and
the results returned from search engines are lacking.


Using these and other opportunistic approaches help ensure that
they complement each other in approximating the publicly available set
of repositories though it does not guarantee the completeness. We
expected that various ways of crowdsourcing the discovery (with incentives to
share a project's Git URL) would help increase the coverage in the
future.

\subsection{Project Retrieval}\label{s:retrieval}

This data retrieval task can be done in
parallel on a very large number of servers but requires a
substantial amount of network bandwidth and storage. The simplest approach
is to create a local copy of the remote repositories via ``git clone''
command. As of May 2019, we estimate over 73M unique
repositories (excluding GitHub repositories marked as forks,
repositories with no content, and private repositories). A single thread shell process on a
typical server CPU (we used Intel E5-2670) with no limitations on
network bandwidth clones between 20K and 50K repositories
(the time varies dramatically with the size of a repository and the
forge) in 24 hours. To clone 73M repositories in one week would,
therefore, require between two and five hundred servers. However, we do not
possess dedicated resources of that size and, therefore, optimize the
retrieval by running multiple threads per server and retrieving a
small subset of the repositories that have changed since the last
retrieval. Specifically, we use five Data Transfer Nodes of a
cluster computing platform which provides 300 nodes in total with a bandwidth up to 56 Gb/s.

\subsection{Data Extraction}\label{s:extraction}

Code changes are organized into commits that typically change one
or more source code files within the project. Once the
repository is cloned as described above, we extract the Git
objects\cite{chacon2014pro}
from each repository. 

\subsubsection{\textbf{Data Model}}
Git~\cite{chacon2014pro} is a content-addressable filesystem
containing four types of objects. The reference to these objects
is a SHA1~\cite{eastlake2001us,wang2005collision,qi2007fast}
calculated based on the content of that object. A few typical Git objects are described below.\\\noindent
\textbf{commit}: A commit is a string including the SHA1's of commit parent(s)
  (if any),  the folder (tree object), author ID and timestamp,
  committer ID and timestamp, and the commit message.\\\noindent
  \textbf{tree}: A tree object is a list
  that contains SHA1's of files (blobs) and subfolders (other trees)
  contained in that folder with their associated mode, type, and name. \\\noindent
\textbf{blob}: A blob is the compressed version of the file
  content (the source code) of a file.\\\noindent
\textbf{tag}: A tag is the string (tag) used to associate readable names with specific versions of the repository. 

\begin{figure}[htbp]
\centering
\includegraphics[width=8cm,height=5cm,keepaspectratio]{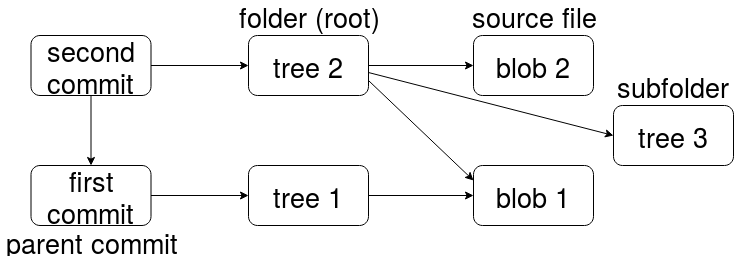}
\caption{Git objects}
\label{fig2}
\vspace{-10pt}
\end{figure}

Figure~\ref{fig2} illustrates the relationships among the Git objects described
above. The snapshot at any entry point (commit) is constructed by
following the arrows from left side to right side. Each commit points to a tree(folder), and each tree points to blobs(files in this folder) inside it and its subtrees(subfolders).

\subsubsection{\textbf{Object Extraction}}
While a standard Git client allows extraction of raw Git objects, it displays
them for manual inspection. For bulk extraction of the Git objects, first
we list all objects within the Git database, categorize them, and then
create a bulk extractor based on a portable pure C implementation of \textit{libgit2}.\footnote{https://libgit2.org/} (libgit2 is a portable, pure C implementation of the Git core methods provided as a re-entrant linkable library with a solid API, allowing you to write native speed custom Git applications in any language which supports C bindings)
We run listing and extraction using 16 threads on each of the 16-CPU node on
a cluster.\footnote{CPU: E5-2670, No. node: 36, No. core: 16, Mem size: 256 GB}
The process takes approximately two hours for a single node to process
50K repositories. 

\subsection{Data Storage}\label{s:storage}
\begin{table}[thb]
\fontsize{9}{8}\selectfont
\centering
\caption{Storage Abstraction}
\label{tab:storage}
\begin{adjustbox}{width=\linewidth}
\begin{tabular}{lllll}
\toprule
\textbf{Abstraction} & \textbf{Schema}  & \textbf{Technology}  & \textbf{\# records (Billion)} & \textbf{size} \\
\midrule
\rule{0pt}{5ex}
Cache/Location Index & \begin{tabular}[c]{@{}l@{}}Key-Value pair\\ Key: Git object SHA1\\ Value: Packed Integer\end{tabular}   & Tokyo Cabinet      & \begin{tabular}[c]{@{}l@{}}Commit: 2.0\\ Tree: 8.3\\ Blob: 7.9\end{tabular} & \begin{tabular}[c]{@{}l@{}}Commit: 121 GB\\ Tree: 487 GB\\ Blob: 487 GB\end{tabular} \\
\hline
\rule{0pt}{5ex}
Value/Content Index  & \begin{tabular}[c]{@{}l@{}}Key-Value pair\\ Key: Git object SHA1\\ Value: offset \& length\end{tabular} & Tokyo Cabinet      & Same as above                                                               & \begin{tabular}[c]{@{}l@{}}Commit: 107 GB\\ Tree: 388 GB\\ Blob: 370 GB\end{tabular} \\
\hline
\rule{0pt}{5ex}
Physical Storage     & \begin{tabular}[c]{@{}l@{}}Concatenated compressed \\ Git object in binary file\end{tabular}            & Binary Compression & Same as above                     &   \begin{tabular}[c]{@{}l@{}}Commit: 542 GB\\ Tree: 8.6 TB\\ Blob: 102 TB\end{tabular} \\
\bottomrule                                
\end{tabular}
\end{adjustbox}
\end{table}

\begin{figure}
    \centering
    \includegraphics[width=\linewidth]{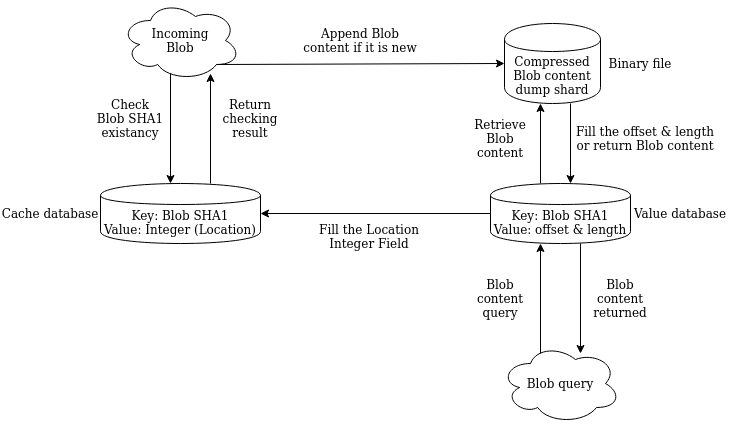}
    \caption{The workflow of storing Git Blob into WoC}
    \label{fig:blob}
\end{figure}

The collection of public Git repositories as a whole replicates many of  the same
Git objects hundreds of times~\cite{M09msr}. Without removing this
redundancy, the required storage for the entire collection exceeds 1.5PB, and
it also makes analytics tasks virtually impossible without extremely
powerful hardware.
Many reasons for this redundancy exist, such as pull-based
development, usage of identical tools or libraries, and copying of
code.

To avoid storing identical Git objects, we store all
Git objects into a single database. The database is organized into
four parts corresponding to each type of Git object. Each part is
further separated into a cache and content.  The cache is used to
rapidly determine if the specific object is already stored in our
database and is necessary for data extraction described above.
Furthermore, the cache helps determine if a specific repository
needs to be cloned or if an object needs to be extracted from the
cloned repository. If the heads (the latest commits in each branch
in .git/refs/heads) of a repository are already in our database,
there is no need to clone the repository altogether.

Cache database is a key-valued database, with the twenty byte Git
object SHA1 being the key and the packed integer (indexing the
location of the object in the corresponding value database) being
the value.  The value database consists of an offset lookup table
that provides the offset and the size of the compressed Git object
in a binary file (containing concatenated compressed Git
objects). The workflow of storing a Git object (e.g. blob) is
  described in Figure\ref{fig:blob}. An incoming blob sha1 is
  checked against the Cache database to see if our system already
  contains it. If not, then we append it to a binary file, then
  generate a new record in the Value database with key being the
  blob SHA1 and value being the offset of the blob in the binary
  file and the length of the blob. A new record is also generated in
  the Cache database with the key being the blob SHA1 and the value
  being a packed Integer pointing out the order of the record (in
  Value database dumped text index file).

While this storage allows for a fast sweep over the entire
database, it is not optimal for random lookups needed, for example,
when calculating diffs associated with each commit. For commits and
trees, therefore, we also create a key value database where a key is SHA1 of the
Git object and a value is the compressed content of the said object.
Cache performance is relatively fast: a single thread on Intel
E5-2623 is capable of querying of 1M Git objects in under 6 seconds,
or over 170K Git objects per second per thread. This can be
multi-threaded and run on multiple hosts, thus reaching any desired
speeds with expanded hardware.

Needless to say, with 18B objects occupying over 120TB we need to use
parallel processing to do virtually anything. Thankfully, we can use
SHA1 itself to split the database into pieces of similar size. We,
therefore, split each of the database into 128 slices based on the
first seven bits of Git object SHA1. This results in 128 key-offset
cache databases for all four types of objects, 128 content databases
as flat files for the four types of objects, and 128 key value
databases for commits and trees: 128*(4+4+2) databases with each
capable of being placed on a separate server to speed up parallel
tasks. The individual databases containing content range from
20MB for tags up to over 0.5TB for blobs. The largest individual
cache databases are over 2Gb for tree object SHA1s.


Databases are fragile and may get corrupted due to
hardware malfunction, internet attack, pollution/loss by
unrecoverable operation, etc. To enhance the robustness and
reliability and to avoid permanent data loss, we maintain three
copies of the databases: two copies on two
separate running servers and one copy on a workstation that is not
permanently connected to Internet. In the future, we will consider
keeping a copy using a commercial cloud service.

Furthermore, due to the size of the data and complexity of the
pipeline, some of the objects may have been missed or have been
retrieved but are not identical to originals. Techniques to validate
the integrity of the data at every stage of the process are necessary. We therefore,
include numerous tests to ensure that only valid data gets
propagated to the next stage.

In particular, the errors when listing and extracting objects are
captured and the operation is repeated in case a problem occurs. The
extracted objects are validated to ensure that they are not corrupt
and also to ensure that they are not going to damage the database or
the analytics layer. To validate correctness, the object is
extracted per Git specifications and recreated from scratch. The SHA1
signature is compared to ensure it matches that of the original
object. A substantial number of historic objects have issues due to
a bug in Git that has since been fixed. Furthermore, a much smaller
number of objects also had issues that we assume are either caused
by problematic implementations of Git or problems in operation
(zero-size objects that may be occasionally created when Git runs out
of disk space during a transaction).

Despite the scrubbing and validation efforts, some of the data may
still be problematic or missing, therefore a continuous process of
checking the database for missing or incorrect data is needed.  We
plan to add a missing object recovery service that identifies missing
commits, blobs, and trees, and retrieves and stores them (in case
they are still available online).

\subsection{Update}\label{s:update}

\begin{figure}[htbp]

\centerline{\includegraphics[width=10cm,height=7cm,keepaspectratio]{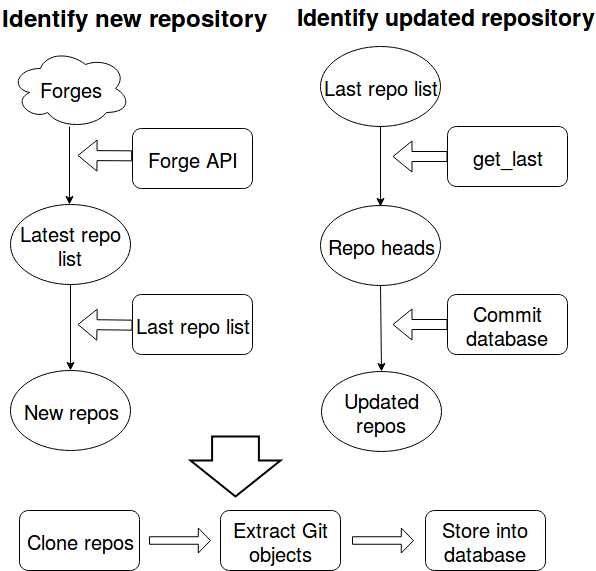}}

\caption{Update workflow}
\label{fig_update}

\end{figure}

The process of cloning all GitHub repositories takes an increasing amount of time with the growth in size of existing repositories and the emergence of new ones,
given fixed hardware. Currently, to clone
all Git repositories (over 90M including forks), we estimate the
total time to require six hundred single-thread servers running for a
week and the result would occupy over 1.5PB of disk space. Fortunately, Git objects are immutable and we can leverage that to simplify and speed up the updates.  More generally, to get acceptable update times,
we use a combination of two approaches:
\begin{itemize}
\item Identify new repositories, clone and extract Git objects
\item Identify updated repository and retrieve only newly added Git
  objects
\end{itemize}
The work flow is illustrated in Fig.~\ref{fig_update}.

In fact, only approximately three million new projects were 
created and an additional two million updated during Dec, 2018.

\subsubsection{\textbf{Procedures for new repositories}}
Forge-specific APIs are utilized to obtain the complete list of public
repositories as described above. A comparison with prior extract yields
new repositories. The list may include renamed repositories and
forks. We can exclude forks for GitHub, since it is an attribute
returned by GitHub API.\footnote{The exclusion only happens when all the conditions are met: a GitHub project has not been seen in the project list in a previous update, GitHub marked this project as a fork, and all the heads already exist in our database.} Other forges contain fewer repositories, so
the forks are not large enough to be a concern.

\subsubsection{\textbf{Procedures for updated repositories}}
First we need to identify updated repositories from the complete list of repositories.
Since we are not sure how GitHub determines
the latest update time for a repository, we use a forge-agnostic way
of identifying updated repositories. 
We modified the
\textit{libgit2} library so that we can directly obtain 
 the latest commit of each branch in a Git
repository 
for an arbitrary Git repository URL, without the need to
clone the repository. If any of the heads contain a commit that
is not already in our database, the repository must have
had updates and needs to be obtained. 

\begin{figure}[htbp]

\centerline{\includegraphics[width=10cm,height=6cm,keepaspectratio]{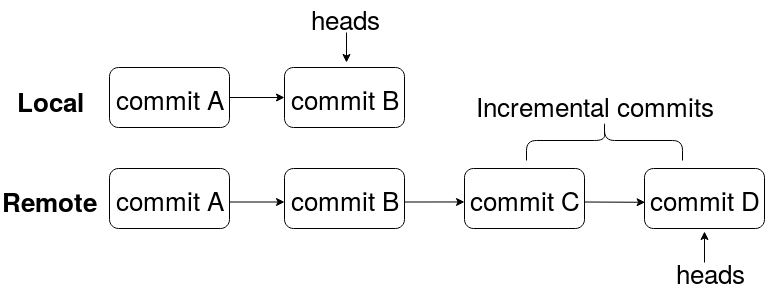}}

\caption{Incremental commits}
\label{Incrementals}

\end{figure}
\begin{figure*}[htbp]
\centerline{\includegraphics[width=12cm,height=1.5cm,keepaspectratio]{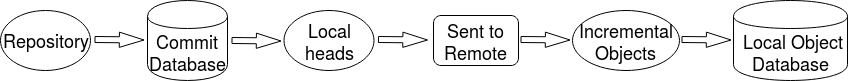}}
\caption{Future workflow}
\label{future_protocol}

\end{figure*}
We are working on a strategy to reduce the amount of bandwidth needed
to do the updates. Instead of cloning an updated repository, we'd like
to retrieve only incremental Git objects (see Fig.~\ref{Incrementals}) that are generated
during the time gap between two consecutive updates.  This can be easily
 done via ``git fetch'' for a Git repository, but since we do not keep
the original Git repository and it is time consuming to pre-populate
it with Git objects, we plan to customize ``git fetch'' protocol by inserting
additional logic in order to use our database backend that comprises
Git objects from all repositories. The procedure  consists of two
steps: 
\begin{enumerate}
\item Customize ``git fetch protocol''\footnote{``git fetch'' downloads only
    new objects from the remote repository} to work without Git's
  native database. 
\item Keep track of the heads for each project that we have in our
  database so that we can identify latest commits to the modified ``git
  fetch''. 
\end{enumerate}

For the second step, the database backend will use the
project name as input and provide the list of heads for the project. These heads are
then sent to the remote so that the set of latest commits (and
related trees/blobs) will be calculated out and transferred back
as illustrated in
Figure~\ref{future_protocol}. By following this strategy, we could
drastically speed-up mining incremental Git objects
from repositories in each update.

\subsection{Data Reorganization}\label{s:transformations}

objects in Git are organized in a way for fast reconstruction of a
repository at each commit/revision. In fact even the seemingly
simple operation of identifying what files changed in a commit is
computationally intensive. Furthermore, there is no consideration for
the projects, files, or authors as first-class objects. This
limits the usability of the Git object store for research and
suggests the need for an alternative data design. Since our
objective is to obtain relationships among projects, developers,
and files, we have created an alternative database that allows both
a rapid lookup of these associations and sweeps through the entire
database that make calculations based on such relationships. 

\subsubsection{\textbf{Analytic Database}}
The scale of the desired database limits our choices. For example, a
graph database\footnote{a database that uses graph structures for
  semantic queries with nodes, edges and properties to represent and
  store data} like neo4j
would be extremely useful for storing and querying relationships,
including transitive relationships. However, it is not capable (at
least on the hardware that we have access to) of handling hundreds
of billions of relationships that exist within the entire FLOSS. In
addition to neo4j, we have experimented with more traditional database
choices.  We evaluated common relational databases MySQL and
PostgreSQL and key value databases or NoSQL~\cite{leavitt2010will}
databases MongoDB, Redis, and Cassandra. SQL like all
centralized databases~\cite{abadi2009data} has limitations handling 
petabyte datasets~\cite{russom2011big}. We, therefore, focus on
NoSQL databases~\cite{moniruzzaman2013nosql} that are designed for large scale
data storage and for massively parallel data processing across a
large number of commodity servers~\cite{moniruzzaman2013nosql}.

For the specific needs of the cache database and for key value
stores for the analytics maps we use a C database library called TokyoCabinet (similar to
berkeley\_db) using a hash-indexed as described above, to provide
approximately ten times faster read query performance than a variety of
common key value databases such as MongoDB or Cassandra. Much faster
speed and extreme portability lead us to use it instead of more
full-featured NoSQL databases.
\subsubsection{\textbf{Maps}}\label{maps}
Apart for the general requirement to be able to represent global
relationships among code, people, and projects, we also consider the basic patterns of data access for several specific research tasks as  use cases in order to design a database
suitable for accomplishing research tasks within a reasonable time
frame. The specific use cases are:
\begin{enumerate}
    \item Software ecosystem research would need the entire set of repositories belonging to a specific FLOSS sub-ecosystem, e.g., the set of all repositories that use Python language.
    \item Developer behavior research would need to identify all projects that a specific developer worked on, the files they
authored, and software technologies they used.
    \item Code reuse research would need to identify all projects where a specific piece of code occurs and determine how it got
there. 
\end{enumerate}


To support the first task, a mapping from file names to project names
would be necessary. The second task would require author to project,
file, and to content of the versions of the file authored by that
developer (in order to access the source code and identify what
components or libraries were employed). The last task would require
a map between blobs (that contain snippets of code) and projects. It
would also require a map between blobs and commits in order to
identify the time when the specific piece of code was introduced.

We have identified a number of objects and attributes of interest
here: projects, commits, blobs, authors, files, and time. The
complete set of possible direct maps for an arbitrary pair is 30.  
Since author and time are properties of the commit
and are not properties of projects, blobs, or files, it makes sense
to place commit at the center of this network database.
The author-to-file map can then be constructed as a composition of
author-to-commit and commit-to-file maps; and author-to-project map
can be constructed via author-to-commit and commit-to-project maps.
We also need to associate file names with the corresponding blobs since 
a single commit may create multiple files. 
Out of the 12 maps,\footnote{bidirectional maps between the commit and  
five objects/attributes and between file and blob} only 10 need to be
instantiated because commit-to-author and commit-to-time maps are
embedded as the properties of the commit object.

In addition to having the commit at the center, for certain tasks
we also needed to have a blob-to-file map as well. For example, we
want to identify module use in Python language files. First, we need
to identify relevant files via suitable extension (e.g., .py), then
we can determine all the associated commits via file to commit map.
These commits, however, may involve other files and if we use commit
to blob map to identify associated blobs, we would get blobs not
just for ``python'', but also for all files that were modified in
commits that touched at least one ``python'' file. The file-to-blob map
allows us to reduce the number of blobs that need to be analyzed
dramatically.  

In addition to these basic maps we create additional maps, such as the author ID to author ID map for 
IDs that have been established to belong to the same person (see Section~\ref{identitymatching}), and 
project to project maps to adjust for the influence of forking. Project-to-project maps are based on the transitive 
closure of the links induced between two projects by a shared commit. Explicit forks that can be obtained as a GitHub project property 
do not generalize to other forges and, even on GitHub, represent only a fraction of all repositories that have been 
cloned from each other and then developed independently. Project-to-project map also handles instances where repositories exist 
on multiple forges or when they are renamed. 

As with the original data we utilize multiple databases and use
compressed files for sweep operations and TokyoCabinet for random
lookup. We separate maps into 32 instead of 128 databases we use for
the raw objects since maps tend to be much smaller in size than, for
example, blobs. For commits and blobs we use the first character of
SHA1 for database identification. For authors, files, and projects,
we use the first byte of FNV-1a
Hash.\footnote{http://www.isthe.com/chongo/tech/comp/fnv/index.html\#FNV-1a} Both
approaches yield quite uniform distribution over bins.

As noted above, the maps from commit to metadata are not difficult
to achieve because most of the metadata is part of the content of a commit
object. However, Git blobs introduced or removed by a commit
are not directly related to the commit and need to be calculated by
recursively traversing trees of the commit and its parent(s). 
A Git commit represents the repository at
the-state-of-world and contains all the trees (folders) and blobs
(files). To calculate the difference between a commit and its parent commit,
i.e., the new blobs, we start individually from the root tree that
is in the commit object, traverse over each subtree and extract each
blob. By comparing two sets of blobs of each commit, we obtain the
new blobs for the child commit. This step requires substantial
computational resources, but the map from the commit to the blobs
authored in a commit is used in numerous research scenarios and,
therefore, is necessary. On average, it takes approximately one
minute to obtain changed files and blobs for 10K commits in a
single thread. With 1.5B commits, the overall time for a single
thread would take 104 days, but it needs to be done only on
approximately 20-40M new commits generated each month. 

\section{Architecture for research workflows}\label{s:tool}

To make WoC more easily usable in a wide variety of research
scenarios, we have designed an architecture to help simplify, support, and evaluate
the implementation of research tasks. This section describes that architecture, along with critical performance benchmarks to inform the 
users on the computational tasks for alternative implementations.

\subsection{Architecture}

\begin{figure}[htbp]
\centerline{\includegraphics[width=0.9\linewidth]{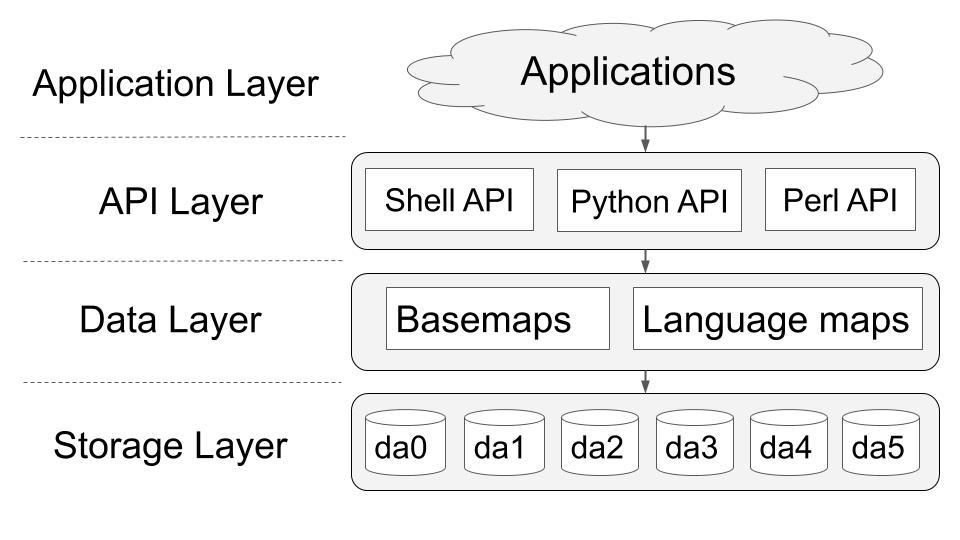}}
\caption{Architecture of the Tool}
\label{f:arch}
\end{figure}

\noindent
The research workflow architecture is illustrated in
Figure~\ref{f:arch}. The figure shows the application layer, built
on top of the three lower layers: \\\noindent
\textbf{Application Layer:} This layer is where the research tasks
are implemented by use of WoC. We provide a library of applications to
illustrates various types of research analyses that can be
implemented using WoC. \\\noindent
\textbf{API Layer:} The applications may use Shell or Python API, or may
reuse or modify Perl files (used to support Shell API) to access and process
the WoC data. A more detailed description of the Shell and Python APIs
can be found in Section~\ref{s:API}.\\\noindent 
\textbf{Data Layer:} 

As described above, to be able to
identify the relationships rapidly we constructed several
types of relationships (or basemaps) that cross-reference the Git objects and other properties. In particular, we treat \textit{project, commit, blob,
  author, file name} as the first class objects and map them to their
properties (e.g., time, parent commit, head commit, child commit,
etc.). In addition to these basemaps, we also construct technical
dependencies that are derived from
importing external dependencies for several languages (Language Maps).
These dependencies are calculated based on each version of a file.
The data is described in more
detail in Section~\ref{ss:data}.\\\noindent 
\textbf{Storage Layer} All the data are hosted on six
servers, which are connected to each other through
NFS (network file system). Users can login to any of the servers (da0
to da5) and and run their applications on multiple servers. 

\subsection{API}\label{s:API}
We support three primary APIs for WoC users to access the
dataset: Shell, Python, and Perl. Presently, running an application
requires being logged in to one of the hosting servers. 
\subsubsection{Shell API}\label{s:shellAPI}

For the lowest level access we provide Shell API that is modeled after 
core philosophy of 
Unix:\footnote{https://en.wikipedia.org/wiki/Unix\_philosophy} 
have a set of specialized commands that are connected 
in a workflow through their standard input and output and via creation of files, E.g., according to  Doug McIlroy: ``Make each program do one thing well. To do a new job, build afresh rather than complicate old programs by adding new features''.
The entire application workflow can be built using exclusively shell and standard Unix utilities such as `join', `sort', `cut',
`uniq', `sed', and `wc' with added specialized commands to extract data from key-value databases. The key information for this API is the knowledge of how to use shell and standard Unix command and the description of the databases. To enable this approach we also provide all databases as key-sorted
(and compressed) text files that can be used with `grep', `join', or
`sort' to produce any desired queries. We also add a random lookup 
operation \texttt{getValue mapname} to access values of a key object in the provided \texttt{mapname}.
In addition, we add the command \texttt{showCnt type} to access the content of each Git object given in the standard input
where \texttt{type} is one of \texttt{tag, tree, commit, blob}.
 A few examples are listed below:
\begin{itemize}
    \item Checking the content of a Git object given a SHA:
    \begin{lstlisting}[language=bash]
    # (on da3) e.g., show a commit SHA's content:
    echo e4af89166a17785c1d741b8b1d5775f3223f510f | showCnt commit
    # Output Formatting: 
    # Commit SHA;Tree SHA;Parent Commit SHA;Author;Committer;Author Time;Commit Time
    e4af89166a17785c1d741b8b1d5775f3223f510f;f1b66dcca490b5c4455af319bc961a34f69c72c2;c19ff598808b181f1ab2383ff0214520cb3ec659;Audris Mockus <audris@utk.edu>;Audris Mockus <audris@utk.edu>;1410029988 -0400;1410029988 -0400
    \end{lstlisting}
    \item Given an object, check its related objects:
    \begin{lstlisting}[language=bash]
    # (on da3) e.g., show the names of the projects associated with a given commit SHA:
    # ``getValue" command takes a database name as an argument and keys presented as standard input and produces key-value pairs as output.
    echo e4af89166a17785c1d741b8b1d5775f3223f510f | getValue /da0_data/basemaps/c2pFullP 
    # Output Formatting: Commit SHA;ProjectNames
    e4af89166a17785c1d741b8b1d5775f3223f510f;W4D3_news;chumekaboom_news;fdac15_news;fdac_syllabus;igorwiese_syllabus;jaredmichaelsmith_news;jking018_news;milanjpatel_news;rroper1_news;tapjdey_news;taurytang_syllabus;tennisjohn21_news
    \end{lstlisting}
\end{itemize}

\subsubsection{Python API}\label{s:pythonAPI}

At the top level of abstraction, we provide Python API via 
package 
\textbf{oscar}\footnote{https://github.com/ssc-oscar/oscar.py} that implements
the key notions of author, file, project, commit, blob, and tree as the corresponding classes.  
The enumeration below describes Python classes that were created by wrapping up data objects~\ref{t:terminology}. Each of the classes has a couple of methods attached to access corresponding properties. For the methods that contain slash(/), the method before slash returns actual data in string, while the one after return a generator of corresponding ``python'' instances.
E.g. \texttt{Author.commit\_shas() }returns a
list of the SHAs of commits that the person authored;
\texttt{Author.commits()} returns a generator of Commit objects
built from those SHAs. 
\begin{enumerate}
    \item Author(`...') - initialized with a combination of name and email, e.g. ``Albert Krawczyk $<$pro-logic@optusnet.com.au$>$"
    \begin{itemize}
        \item .commit\_shas/commits - all commits by this author
        \item .project\_names - all projects this author has committed to
    \end{itemize}
    \item Blob(`...') - initialized with SHA of blob
    \begin{itemize}
        \item .commit\_shas/commits - commits creating or modifying (but not removing) this blob
    \end{itemize}
    \item Commit(`...') - initialized with SHA of commit
    \begin{itemize}
        \item .blob\_shas/blobs - all blobs in the commit
        \item .child\_shas/children - the commit that follows this commit
        \item .changed\_file\_names/files\_changed
        \item .parent\_shas/parents - the commit that this commit follows
        \item .project\_names/projects - projects this commit appears in
    \end{itemize}
    \item Commit\_info(`...') - initialized like Commit()
    \begin{itemize}
        \item .head
        \item .time\_author - the commit time and its author
    \end{itemize}
    \item File(`...') - initialized with a path, starting from a commit root tree. This represents a filename, regardless of content or repository; e.g. File(``.gitignore") represents all .gitignore files in all repositories.
    \begin{itemize}
        \item .commit\_shas/commits - All commits that include a file with this name
    \end{itemize}
    \item Project(`...') - initialized with project name/URI
    \begin{itemize}
        \item .author\_names - all author names in this project
        \item .commit\_shas/commits - all commits in this project
    \end{itemize}
\end{enumerate}

\subsubsection{Perl APIs}\label{s:perlAPI}

While the Python API provides high level of abstraction, it is not very 
computationally efficient. In order to provide an intermediate level 
of efficiency between that of Python and Shell APIs, we also provide 
a way to implement applications or their components in Perl language.
For example, the shell commands \texttt{getValue} and \texttt{showCnt} are both implemented in Perl. 
The Perl API instead of
creating classes of objects as in Python, it handles the
maps directly. To support writing WoC workflows in Perl we provide 
a variety of utility functions
in package `WoC.pm.' We also have, over the course of evolving WoC, created a number of applications that can be used as
templates and modified by the users for their needs.
For example, we can parse the content of the commit to 
obtain its tree, parent commit, author, and time:
\begin{lstlisting}[language=perl]
use WoC; 
my ($tree, $parent, $authName, $authEmail) = ("","","","");
my ($pre, @rest) = split(/\n\n/, $code, -1);
for my $l (split(/\n/, $pre, -1)){
  $tree = $1 if ($l =~ m/^tree (.*)$/);
  $parent .= ":$1" if ($l =~ m/^parent (.*)$/);
  ($authName, $authEmail) = gitSignatureParse($1) if ($l =~ m/^author (.*)$/);
}
($auth, $ta) = ($1, $2) if ($auth =~m/^(.*)\s(-?[0-9]+\s+[\+\-]*\d+)$/);
$parent =~ s/^:// if defined $parent;
\end{lstlisting}

We also have examples on how to parse, for example, a snippet of ``python'' 
source code to obtain the 
dependencies defined by the import statements. A ``segment'' is shown below: 
\begin{lstlisting}[language=perl]
for my $l (split(/\n/, $code, -1)){
 if ($l =~ m/^\s*import\s+(.*)/) {
  my $rest = $1;
  $rest =~ s/\s+as\s+.*//;
  my @mds = $rest =~ m/(\w[\w.]*[\,\s]*)*/;
  for my $m (@mds) { $matches{$m}++ if defined $m};
 }
 if ($l =~ m/^\s*from\s+(\w[\w.]*)\s+import\s+(\w*)/) {
  if ($2 ne ""){ $matches{"$1.$2"} = 1; }
  else{ $matches{$1} = 1; }
 }
}
\end{lstlisting}
For more detail please refer to the tutorial page of our repository.\footnote{https://bitbucket.org/swsc/lookup/src/master/}

\subsection{Description of the WoC Data}\label{ss:data}
\begin{figure}[htbp]
\centerline{\includegraphics[width=0.9\linewidth]{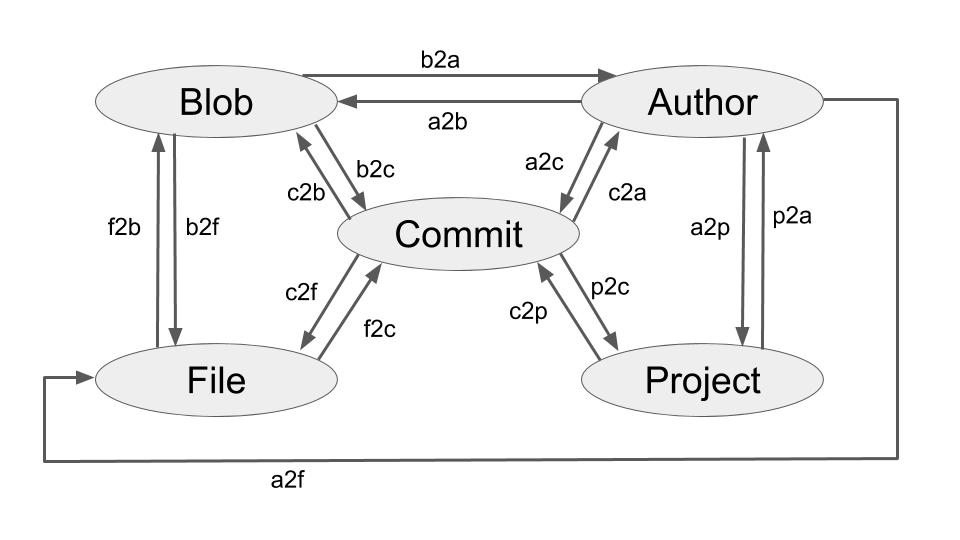}}
\caption[Maps between primary objects (basemaps)]{Maps between primary objects\protect\footnotemark (Basemaps)}
\label{f:Tapajit}
\end{figure}

We use abbreviated object names for WoC data and basemaps as shown in
Table~\ref{t:terminology}. As noted above, types of basemaps are created to
represent relationships among these objects, which are illustrated
in Figure~\ref{f:Tapajit}. Notice that some maps are missing in
Figure~\ref{f:Tapajit}, because initially we built maps with commit
being the core, and other maps were built as certain research tasks
the users were attempting to do would benefit from them. 
The basemaps are stored in TokyoCabinet databases for random queries
and key-sorted compressed text files of these basemaps are
also created to enable quick sweeps over the whole dataset and to
enable the shell API.   

In addition to the basemaps, programming language based maps are created
to enable language oriented analytic and applications. These contain
mappings that list repositories, and the modules they depended on,
at a given UNIX timestamp under a specific commit. The format of
each entry in these maps are like the following, where
\verb|module1;module2;...| represent the modules that repository
depended on at the time of that commit: 
\begin{verbatim}
    commit;repository_name;timestamp;author;blob;module1;module2;...
\end{verbatim}
At the time of writing, 12 maps are ready including C, C\#, Java,
JavaScript, Python, R, Rust, Go, Swift, Scala, and Fortran with more
language maps anticipated to be added in the future.

\begin{table}[thb]
\fontsize{9}{8}\selectfont
\centering
\caption{Naming Conventions}
\label{t:terminology}
\begin{adjustbox}{width=0.5\linewidth}
\begin{tabular}{ l l l }
\toprule
\textbf{Object Abbreviation} & \textbf{Annotation} & \textbf{Entity Type} \\
\midrule
a    &  author      & string     \\
b    &  blob        & SHA   \\ 
c    &  commit      & SHA     \\ 
f    &  file name       & string   \\ 
p    &  project     & string      \\ 
\bottomrule
\end{tabular}
\end{adjustbox}
\end{table}

\footnotetext{`File' in this figure refers to `File name'}

\begin{table}[thb]
\fontsize{9}{8}\selectfont
\centering
\caption{Maps Size}
\label{t:maps_size}
\begin{adjustbox}{width=\linewidth}
\begin{tabular}{ l l l l }
\toprule
\textbf{Map Type} & \textbf{Map Size (GB)}  & \textbf{Text Dump Size (GB)} & \textbf{\# of Records (Billion)} \\
\midrule
a2b      & 311       & 232                 & 9.9          \\
a2c      & 169       & 50                  & 2.0         \\
a2p      & 44        & 28                  & 3.7          \\
a2f      & 282       & 191                 & 31.2         \\
b2a      & 1102      & 679                 & 9.9         \\
b2c      & 1317      & 1152                & 40.7         \\
b2f      & 1099      & 588                 & 35.2         \\
c2a      & 192       & 91                  & 2.0          \\
c2b      & 970       & 998                 & 40.7         \\
c2f      & 695       & 449                 & 60.1         \\
c2p      & 1127      & 797                 & 101.2        \\
f2a      & 1955      & 380                 & 31.2         \\
f2b      & 1574      & 1085                & 35.2         \\
f2c      & 2393      & 1176                & 60.1         \\
p2a      & 84        & 52                  & 3.7          \\
p2c      & 1982      & 2283                & 101.2   \\
\bottomrule
\end{tabular}
\end{adjustbox}
\end{table}

\subsection{Performance Benchmark}

The anticipated workflow of a specific research task involves a set
of queries that proceed from selecting an initial sample of interest
such as a set of files related to a specific language, a set of
projects or authors with certain properties or other collection. This
is typically followed up by one or more network operation such as
identifying blobs associated with the selected files, projects
associated with the initial set of developers and so on. These tasks can typically be implemented in numerous ways, each leading to
different computer memory, disk IO, and computational overheads. To
help users decide upon the the best way to proceed and, more
generally, to gauge the time needed for their desired workflow, we
set up experiments to test our WoC infrastructure performance on
such queries. Our existing basemaps should meet users' need in most
cases by a query of a single map (e.g. author to commit). However,
in cases where a map is not ready (e.g. file to project in
Figure~\ref{f:Tapajit}), users might need to combine/join two or
more maps to achieve their goal. We, therefore, tested the performance of both single
map queries and combined map queries, and present the results below.

\begin{figure}[htbp]
\centerline{\includegraphics[width=0.9\linewidth]{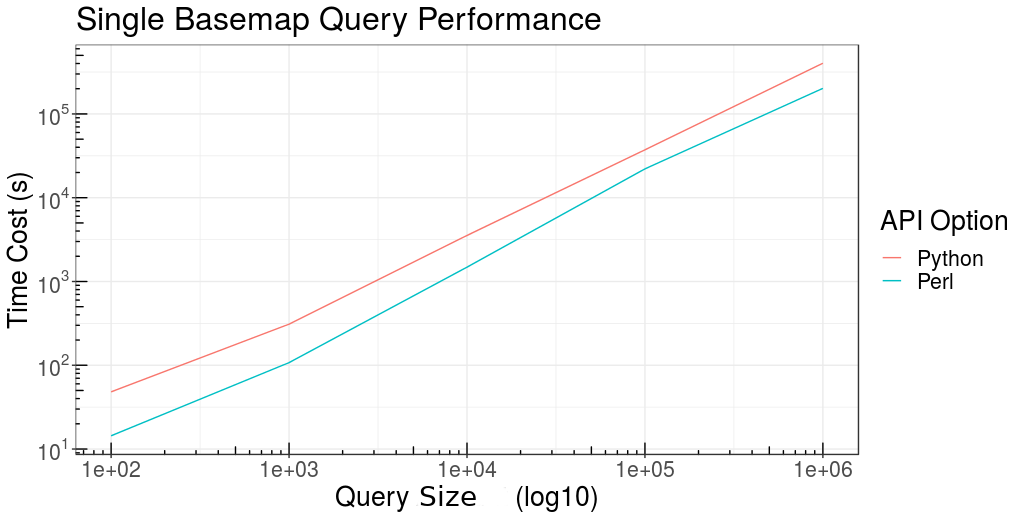}}
\caption{Single Map Query Performance}
\label{f:Single}
\end{figure}

\begin{figure}[htbp]
\centerline{\includegraphics[width=0.9\linewidth]{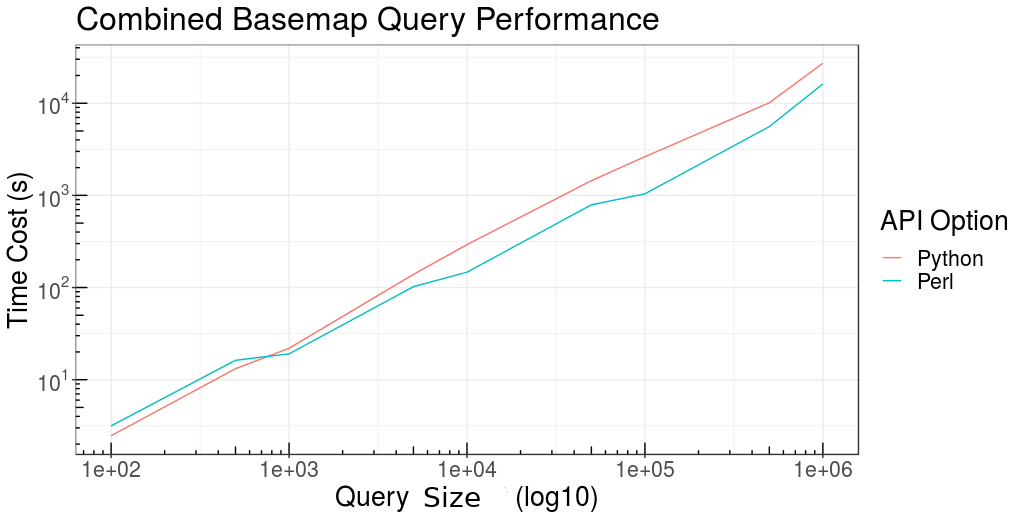}}
\caption{Combined Maps Query Performance}
\label{f:Combined}
\end{figure}

Since the file\footnote{By file, we refer to the file name
  (including folder path) in the rest
  of our paper.} to project map is not pre-computed, we can start from the file
to commit map to test single map query performance and then join the
results with the 
commit to project map to test the combined map query. 
We randomly selected
100, 1K, 10K, 100K, and 1M file names from our dataset, and used the
Python and Shell APIs without any other task being run on the server
to find the corresponding commits in which the
files were modified and the projects those commits belong to. We
collected the time it took to run each test and show them in
Figure~\ref{f:Single} for the single map queries, and
Figure~\ref{f:Combined} for the combined map queries. 

From Figures~\ref{f:Single} and~\ref{f:Combined}, we see that the
run time increases linearly as the task size increased,
highlighting the scalability of the WoC infrastructure. We
also found Shell API to be three to four times faster than Python
API (Figure~\ref{f:Single} and right part of
Figure~\ref{f:Combined}), for the same query. One
hypothesis is the interpreted nature of Python.
Specifically, the data access parts of Shell API are implemented in
Perl. While Perl is an interpreted language just as Python, many of
the functions are implemented natively in C language, while in
Python more performance-critical code is interpreted. 

It is worth noting that the x-axis on Figure~\ref{f:Combined} represents
the number of queries, which in this scenario is the sum of the
number of file to commit queries and the number of commit to project
queries.  

We tested the performance of the tool for 100 to 1M queries. If a
research workflow involves the initial sample of objects for a very
large part of the WoC database,  we recommend leveraging the
database in the form of compressed text for key-value basemaps instead,
because as the number of random access queries increases, it exceeds
the time it takes to sweep the entire database using efficient
shell commands such as grep. 
In fact, a single sweep of the file to commit compressed data only
takes 38 hours while 1M queries of
the file to commit basemap takes 56 hours using Shell API. 

\section{Applications}\label{s:applications}

To evaluate if the experimental platform is capable of supporting
research tasks conducted as a part of actual investigations and 
to provide a set of vignettes for other researchers, we
conducted two types of studies. First, we implemented several basic and
involved research tasks that require the entirety of FLOSS data as a
part of the investigation. Furthermore, we also recruited three
researchers external to our group to either conduct investigations of their
own utilizing WoC or to provide us with their research problems 
that can only be solved by using WoC. 
Below we report both the experiences and results from
these experiments.

\subsection{Use of programming languages}\label{ss:pl}

Language popularity may influence developers decisions as it may affect the market for their software as well as their job prospects. For example:  
What language-specific API should developer provide for their component?
What language should the developer use to implement their product? 

\begin{figure}[htbp]

\centerline{\includegraphics[width=12cm,height=8cm,keepaspectratio]{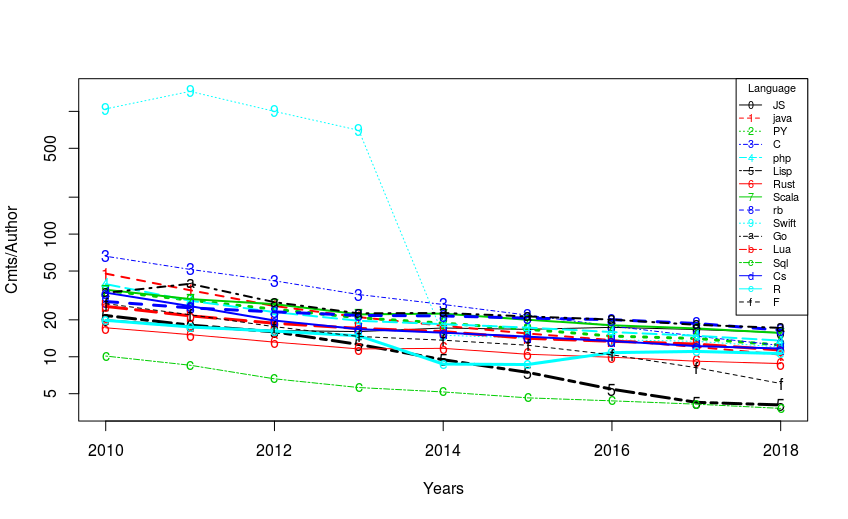}}
\caption{Productivity by Language}
\label{Cmt/Auth}
\end{figure}

To plot, for example, Java language use trend we use WoC to identify all files 
with .java extension. Then, via file-to-commit map, obtain the complete set of commits
authoring these files. Commit dates are 
used to plot the time trends of language-specific commits, authors 
(property of a commit), projects (via commit to project map) and, 
if desired, lines of code changed.
The entire process is highly parallelizable since each map is separated into 32 instances and can be processed independently.
The entire calculation, while not interactive on our hardware, can be performed in tens of minutes. For illustration, 
we show the ratio of the number of
commits over the number of developers (a measure of productivity) each year in Fig.~\ref{Cmt/Auth}.  The ratio decreases for most languages, perhaps because as a language becomes more popular, the
less experienced contributors join and lower the average productivity.

\subsection{Correcting Developer Identity Errors}\label{identitymatching}

One of the particularly troubling data quality issues with version control systems is developer name disambiguation. Often, names and emails of developers are  missing, incomplete, misspelled or duplicate~\cite{german2003automating,Bird:2006:MES:1137983.1138016}.
Performance of any disambiguation algorithm depends on the distribution of the actual misspellings in the underlying data. In order to design and evaluate corrective algorithms, it is important to study a large collection of actual data and unearth patterns of irregularities that compromise data quality. WoC contains a nearly complete collection of Git author ids (name and email combinations) and is, thus, more representative of such irregularities than any specific project. 

To obtain author IDs we use author-to-commit map containing roughly 30 million distinct author IDs. Common error patterns include organizational ids and emails (Mozilla, Linux, Google etc), names of tools and projects (OpenStack, Jenkins, Travis CI), roles such as (admin, guest, root etc.) and words that preserve anonymity (student, nobody, anonymous etc) as a part of their credentials. We also found a large number developer IDs to be misspelled.  

Traditional identity correction approaches rely on the misspelling patterns of author ID (the full name and email)~\cite{Bird:2006:MES:1137983.1138016,Winkler,Winkler06overviewof}. 
With WoC data, we can enhance the traditional string matching with behavioural comparison, by creating similarity measures between author IDs using files modified by developers, time patterns of commits, and writing styles in commit messages. For illustration --- two author IDs that modify a similar set of files may suggest that these IDs belong to the same developer. To implement file-based similarity, we  
used author to commit and commit to file maps to obtain the set of files modified by a single author ID. Then file-to-commit and commit-to-author maps were used 
to calculate similarity using weighted Jaccard measure. Commit message text was used to fit a Doc2Vec~\cite{doc2vec} model to associate each author ID with their writing style. Traditional and behavioural similarities were used to train highly accurate machine-learning model~\cite{alfaa}.


 
This experiment demonstrates the utility of WoC data for designing tools to solve common and vexing data quality problems when constructing developer networks. 
It is also an example of how WoC can be enhanced by incorporating such techniques and providing corrected data to researchers.

\subsection{Cross-ecosystem comparison studies}\label{s:eco}

A second research group used the database to gather comparative statistics about different software ecosystems.  In that research, ``ecosystem'' was defined as the set of packages provided by a (usually single language) package management system such as CRAN (for the R language) or npm (for Node.js packages). The purpose was to supplement other comparative data about such package ecosystems in support of a study of how ecosystem tools and practices influence development behavior.  The ecosystem study involved a survey, interviews, and data mining over 18 ecosystems whose repositories listed more than 1.2M packages.  Some questions about ecosystem practices could be mined from metadata available elsewhere; for example detailed information about dependencies, release frequency, and version numbering practices can be easily extracted from libraries.io.\footnote{https://libraries.io/}  However deeper questions about project content would have been out of reach without WoC;
independently building the mechanism to collect all of these projects, building a database of blobs, files, projects, and authors, and comparing them using various metrics would have been too much work for too little gain without the availability of this research platform.

\subsubsection{\textbf{File cloning across ecosystems}} One such statistic is rate of file cloning. It was theorized that in ecosystems with more flexible support for dependencies and a tolerance for the risk of breaking changes, developers would be more likely to use dependency management tools to make use of functionality from other projects, rather than copying those files in directly; hence in such ecosystems we should find relatively few commits adding a blob that already exists in any other project available through the ecosystem's dependency management system. 

Using WoC, this analysis was straightforwardly accomplished by joining blob-to-commit and commit-to-project mappings, filtering for blobs that appeared in multiple projects, and identifying pairs with one commit in the time frame, and at least one older commit.  Such blobs were discarded when the files were very small (since these often turned out to be empty or trivial files duplicated by chance or by tools) resulting in a set of duplicates that, on visual inspection of a sample, did appear to represent genuine examples of reuse-by-cloning.

\begin{figure}[htbp]

\centerline{\includegraphics[width=10cm,height=6cm,keepaspectratio]{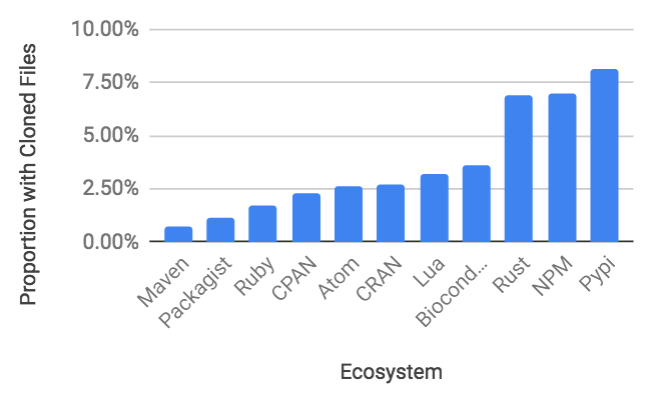}}

\caption{Proportion of repository packages that added at least one cloned code file over 1kb in 2016.}
\label{ClonesByEco}

\end{figure}

Contrary to our expectations, the ecosystem with the most propensity for cloning was the one with the modern and flexible dependency system: npm.  Despite the strengths of npm's dependency management system, there is a strong tradition of copying dependencies like jQuery into projects rather than letting npm retrieve them.  Figure~\ref{ClonesByEco} summarizes the findings for a selection of ecosystems.
\subsubsection{\textbf{Developer migration across ecosystems}} Another metric of interest was developer overlap between ecosystems.  Our ecosystem comparison had included a survey of values and practices in the 18 ecosystems of interest, and we hypothesized that ecosystems might be similar if many developers were actually working in both ecosystems, or had migrated from one to the other.

This question was answered by joining author-to-commit and commit-to-project data for the 1.2M projects in our study, and relying on the identity matching technique described in Sec~\ref{identitymatching}.

Over all pairs of ecosystems, we found a sizable correlation between similarity of average responses on ecosystem \emph{practice} questions (things like frequency of updating, collaboration with other projects, means of finding out about breaking changes), 
and overlap in committers to those ecosystems (Spearman $\rho=0.341, p<.00001, n=16$ ecosystems).
Interestingly, perceived \emph{values} of the
ecosystem (such as a preference for stability, innovation, or replicability) do \emph{not} seem to align with developer overlap ($\rho= -0.05, p=0.44$).  
While more research is needed, we hypothesize that developers may carry practices over from other languages and platforms they have used in the past, in a sometimes cargo-cult-like way, despite recognizing that a new ecosystem is designed to accomplish different ends.

In our very large-scale, wide-ranging study, these questions of developer migration and cloning were of great interest, but would likely have been too expensive to pursue alongside other lower-hanging fruit, absent WoC's prepared set of precomputed maps between files, blobs, authors, projects, and timestamps.  The dataset with its analytical maps was not designed with these particular ecosystem comparison in mind, but its design happens to make such ecosystem questions relatively easy to answer.  

\subsection{Python ecosystem analysis}\label{s:py}

An external researcher wanted to use WoC to investigate open source sustainability by identifying source code repositories for packages in PyPI ecosystem and to 
measure package usage directly.
While over 90\% of npm packages provide repository URLs, less than 65\% of Python Package Index (PyPI) packages do.


The researcher obtained all packages from PyPi and calculated blob SHA1s for \textit{setup.py} file
of the first PyPI releases of each package. 
We filter out resulting 101584 blobs to exclude empty or uninformative blobs (blobs that appear in more than one commit using blob-to-commit map). 
The 54218 informative blobs are then mapped to 54062 unique commits and 
commits to 51924 unique projects (adjusted for forking as described in Section~\ref{s:transformations}). Repositories were recovered 
for 96\% of the 54218 original packages in approximately 20 minutes of computation. To ensure that these repositories are, in fact, used to version control corresponding packages, they can be matched via additional blobs for \textit{setup.py} and other files obtained from PyPi for that package.
Another problem being solved by this researcher was identifying which of the seemingly abandoned projects may be ``feature complete,'' i.e. 
already have the intended scope and do not require further maintenance~\cite{Coelho2018}. 
Feature complete projects should be widely  used in contrast to abandoned projects. Proxies of project usage, e.g., GitHub stars or forks can be used to identify such projects~\cite{Coelho2018}.
WoC, however, lets us measure the extent of use directly. As described in Section~\ref{ss:pl}, all commits modifying Python files 
are identified (file-to-commit map) and the resulting commits are mapped to projects (commit-to-project map). 
Blobs associated with these commits (commit-to-blob map) are then used to extract imports from these files.
The entire procedure could be completed in approximately four hours using the parallelism of the analytic maps (32 databases) and 
blob content maps (128 databases). 


The reported usage was compared to project development activity,
i.e\ the total number of adoptions versus the total number of commits.
In some cases, usage was not accurately reflected in the number of commits.
Common examples are packages providing console scripts and CMS-like projects.
In the former case, packages are not reused in programmatic code and thus don't get into statistics.
In the latter case, website builders often do not publish their code and thus such usage remains unobserved.
Therefore, while the number of public reuses provides some extra information about package use,
it should be adjusted for package type.

\subsection{Repository filtering tool}

Millions of repositories on GitHub and other forges also include projects that are 
completely unrelated to software development. 
GitHub is widely used for education and other 
tasks such as backing up text files, images, or other data. Researchers investigating education may need to focus on tutorials, while other 
researchers may need a sample of actual software development projects. 
Furthermore, a way to select specific subsets of software development projects in order to conduct, for example, ''natural experiments'' would also be highly beneficial. 
WoC can support such project segmentation tasks in a variety of ways. An external education researcher wanted to understand the impact of self-administered programming tutorials. To do that, WoC was used to identify developers who participated in tutorials by searching the set of projects in WoC via keywords related to education: ``assignment'', ``course'', ``homework'', ``class'', ``lesson'', ``tutorial'', ``syllabus'',  ``mooc'', ``udacity''. The search yields over 1M projects. While it is only a small fraction of all projects in WoC but it represents a large sample in absolute terms.  Further filtering was needed to find developers who also worked on actual software projects to measure the impact of self-administered tutorials. The project-to-commit map identified 605K users of tutorials and, when these users were mapped to all projects they participated in, we determine that only half of them contribute to non-tutorial projects. These 300K individuals are potential subjects of tutorial-impact study. Further information (such as their commit activity and project participation) can be obtained from WoC and combined other data, be used in this research. 
WoC can be extended with other approaches to segment projects\footnote{Section~\ref{identitymatching} shows how WoC can also be used to improve them}. For example,  
identification of projects with sound software engineering practices~\cite{munaiah2017curating} relies on a combination of factors easily obtainable in WoC, such as history, license, and unit tests.

\subsection{Other Applications}
A number of research publications have utilized the WoC database, including:
\begin{itemize}
    \item The relationship between dependencies of NPM packages,
      collected using the WoC infrastructure, and their popularity
      was discussed in~\cite{dey2018software}. A related work exploring the interrelationship between software quality and popularity was discussed in~\cite{dey2020deriving,dey2018modeling}
    \item The effort contribution and demand patterns of the
      contributors to the NPM ecosystem was discussed
      in~\cite{dey2019patterns}. 
    \item The investigation of what attributes drive the adoption of a software technology was discussed in~\cite{ma2019methodology}. 
    \item The effect of overall expertise of software developers, extracted using WoC data, and other social and technical factors on the chance of their pull requests getting accepted was discussed in ~\cite{dey2020effect,dey2020pull}, and the related dataset was made available at~\cite{dey2020dataset}.
    \item A method of representing the medium-granularity expertise of developers using a ``skill-space'' based on the APIs they use, and its usefulness in addressing a number of important SE research questions was explored in~\cite{dey2020representation}.
\end{itemize}

\section{Archetypal Usage of WoC}\label{s:tutorial}

To increase the utility of this project to a wider research community, we would like to prioritize easy
access to the World of Code to other interested parties.  In this section,
we provide a brief introduction and an overview of the World of Code
and how to use it. Moreover, there are some resources already in
place that were designed to assist in this process, which can be
found in a public repository.\footnote{
  https://github.com/ssc-oscar/lookup}  
After describing WoC and its applications, in this section 
we demonstrate how to actually use WoC to implement a specific
analysis. A couple of approaches presented here
leverage the WoC tool to implement the Java language trend
analysis, as described in Section~\ref{ss:pl}.  

\begin{enumerate}
    \item Identify Java files based on `.java' extension, collect
      commits that changed these files, and deduplicate the
      commits. Now we have all commits where one or more java files
      were created/modified. The source code of the custom
      \texttt{lsort} command is presented in Appendix~\ref{lsort}. 
    \begin{lstlisting}[language=bash]
    #start from basemap dump("file to commit" dump, P represents version), 
    for i in {0..31}; do zcat /da0_data/basemaps/gz/f2cFullP.$i.s | awk -F ";" "/.java;/{print $2 }" done | ~audris/bin/lsort 10G -u | gzip >JavaCommits.gz 
    \end{lstlisting}
    \item For each commit in commit collection, we can use either
      Python or Perl API to find related author and commit time, and
      then calculate the number of authors and commits by year --
      the trend 
    \begin{lstlisting}[language=python]
    # Using Python
    import gzip
    from datetime import datetime
    from collections import defaultdict
    
    year2commit_count = {}
    year2commit_count = defaultdict(lambda: 0, year2commit_count)
    year2author_count = defaultdict(set)
    java_commits = gzip.open("JavaCommits.gz", "r")
    for commit in java_commits:
        time, author = Commit_info(commit).time_author
        year = datetime.fromtimestamp(int(time)).year
        year2commit_count[year] += 1
        year2author_count[year].add(author)
    print(year2commit_count)
    for year, authors in year2author_count.items():
        print("Year: "+ str(year) + "# of authors: " + str(len(authors)))
    \end{lstlisting}
    \begin{lstlisting}[language=perl,  escapeinside={\%*}{*)}]
    # Using Perl
    # we can run /da3_data/lookup/showCmt.perl on every commit and extract author and time info from there
    # A simpler way is to utilize basemap c2taFullP.{0..31}.tch (i.e., the basemap from commit to author and commit time) by calling Cmt2ATShow.perl (see source code in Appendix)
    zcat JavaCommits.gz | perl Cmt2ATShow.perl | gzip > JavaYearAuthor.gz
    # count records for each year, we get the number of commits by year. E.g., for year 2014:
    zcat JavaYearAuthor.gz | grep "^2014;" | wc -l
    # after deduplication, count records for each year and we get the number of authors by year. E.g., for year 2014:
    zcat JavaYearAuthor.gz | sort -u | grep "^2014;" | wc -l
    \end{lstlisting}
\end{enumerate}

\noindent In fact, directly using language maps is more efficient when implementing this analysis, since language specific information have already been extracted from base maps and stored as language maps for use. 

\begin{lstlisting}[language=perl]
# Alternatively, we use language map: c2bPtaPkgPjava, which consists of commit, blob, project name, time, author, etc. 
zcat c2bPtaPkgPjava.{0..31}.gz | cut -d\; -f3,4 | gzip > JavaYearAuthor.gz
# now follow the similar approach in Perl example shown above to get the final result
\end{lstlisting}

\section{Platform Comparison}\label{s:platform}

\begin{table}[]
\begin{tabular}{lllll}
                                               & \textbf{WoC} & \textbf{SH} & \textbf{GHA} & \textbf{GHT} \\
\textbf{Metadata}                              &              &             &              &              \\ \hline
\textbf{commit}                                & yes          & yes         & yes          & yes          \\
\textbf{authors}                               & yes          & yes         & yes*         & yes*         \\
\textbf{filenames}                             & yes          & yes         & no           & no           \\
\textbf{trees}                                 & yes          & yes         & no           & no           \\
\textbf{blob SHA}                              & yes          & yes         & no           & no           \\
\textbf{Issues, PRs, comments}                 & no           & no          & yes          & yes          \\
\textbf{branches}                              & no           & yes         & no           & no           \\
\textbf{repositories}                          & yes          & yes         & yes          & yes          \\
\textbf{}                                      &              &             &              &              \\
\textbf{Indexing}                              &              &             &              &              \\ \hline
\textbf{authors}                               & yes          & yes         & no           & yes          \\
\textbf{commits}                               & yes          & yes         & no           & yes          \\
\textbf{filenames}                             & yes          & yes         & no           & no           \\
\textbf{blobs}                                 & yes          & yes         & no           & no           \\
\textbf{repositories}                          & yes          & yes         & no           & yes          \\
\textbf{commit to parent}                      & yes          & yes         & no           & yes          \\
\textbf{commit to filenames}                   & yes          & yes         & no           & no           \\
\textbf{commit to head}                        & yes          & no           & no           & no           \\
\textbf{commit to root}                        & yes          & no           & no           & no           \\
\textbf{blob to first commit/author}           & yes          & no          & no           & no           \\
\textbf{time of the commit}                    & yes          & yes         & yes          & yes          \\
\textbf{}                                      &              &             &              &              \\
\textbf{Other features}                        &              &             &              &              \\ \hline
\textbf{language use}                          & per blob     & no          & per repo     & per repo     \\
\textbf{Blob accessible for analysis en masse} & yes          & no          & no           & no           \\
\textbf{Blob contents}                         & yes          & yes         & no           & no           \\
\textbf{Deforking}                             & yes          & no          & no           & no           \\
\textbf{Merge user identities}                 & yes          & no          & no           & no           \\
\textbf{}                                      &              &             &              &             
\end{tabular}
\caption{Features for comparable repositories: WoC= World of Code; SH = Software Heritage; GHA = GithubArchive; GHT = GHTorrent.  `*': GithubArchive and GHTorrent only collect github identities as authorship information. The `Metadata' section describes information that is captured by the platform; `Indexing' describes what indexes are provided for efficiently making connections.  For example although it is possible in Software Heritage to iteratively trace a chain of commits to the chain's root, there is not a table directly connecting them; thus ``commit to root'' has ``no'' in column SH.}
\label{t:platform_feature_comparison}
\end{table}

In this section, we compare different platforms and datasets to
better highlight the unique features of WoC. In particular, we
provide the basic size comparisons\footnote{The size information for
  Software Heritage, GHTorrent and WoC is directly obtained from
  their official websites. GHArchive, on the other hand did not
  provide such detailed information, and we looked into its dataset
  and checked the author ID field and project ID field.}  (see
Table~\ref{t:platform_dataset}) and data categories (see
Table~\ref{t:platform_feature_comparison}) for Software Heritage,
GHTorrent, WoC, BOA, and GHArchive.

Table~\ref{t:platform_dataset} shows that WoC is comparable 
in size of various data components to Software Heritage, GHArchive, and
GHTorrent. All four are much larger than BOA dataset.

Because the platforms have different goals, they collect and index
different kinds of data; Table~\ref{t:platform_feature_comparison}
summarizes some of the differences in data content, indexing, and
services.  While WoC attempts to be comprehensive in capturing
source code and source code change history and authorship, it does
not capture some of the social interaction associated with open
source software engineering, such as bug reports, patch submissions,
and code reviews: GitHub enables these interactions, and they are
captured by pure Github archives like GHArchive and GHTorrent.
WoC, on the other hand, provides more comprehensive cross-indexing
of software artifacts than the other platforms do, and provides
services for the common research steps of merging duplicate
repositories and duplicate user identities.  WoC and Software
Heritage also capture the actual content of software files, which
GHArchive and GHTorrent do not attempt to do, to keep their data
set sizes tractable.  WoC manages the size by provisioning a very
large storage space on a set of linux servers and giving researchers
command-line accounts on these servers. Software Heritage in
contrast provides an open API for querying file contents one at a
time, which makes access easier for a wider group of researchers,
but at the cost of making it difficult for outside researchers to
run analyses over large numbers of source files. Note that BOA is
omitted from this table, since BOA is a tool that could, in theory,
be applied to the task of capturing any of this data, and indexing
it in any of these ways.

\begin{figure}[htbp]
\centerline{\includegraphics[width=12cm,height=6cm,keepaspectratio]{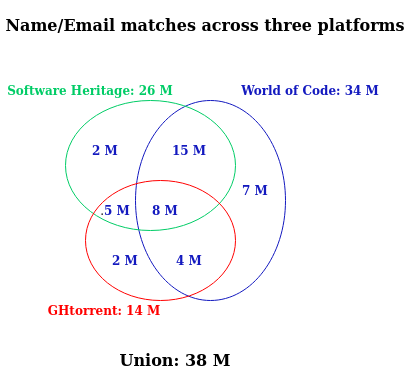}}
\caption{Developers ID match across WoC, Software Heritage and GHTorrent platforms.}
\label{ID-match}
\end{figure}

Moreover, we conduct a performance comparison for specific tasks
across BOA and WoC, and report the result in
Section~\ref{ss:BOA}. In order to understand the data completeness
of different datasets, we compared the corresponding author Id (name
and email) dataset for Software Heritage, GHTorrent and WoC, and
present the result in Figure \ref{ID-match}. We found that over a
total number of 38 million author IDs, WoC captured a majority part
(34 million), which is much larger than the other two datasets (26
million and 14 million). Note that this comparison was conducted in
November 2019, which does not represent the current status. Also,
GHArchive and ghTorrent also capture users who have not made any
code changes, hence such users will not be present in WoC or
Software Heritage datasets that deal exclusively with the source
code and its changes.

In below, we briefly describe the features of each platform, and provide functionalities comparison with WoC.
\begin{table}[thb]
\fontsize{9}{8}\selectfont
\centering
\caption{Dataset Comparison}
\label{t:platform_dataset}
\resizebox{\linewidth}{!}{
\begin{tabular}{ p{3cm} p{2cm} p{1cm} p{1cm} p{1.2cm} p{1.5cm} p{1.5cm} p{1.2cm} p{1.5cm} }
\toprule
\textbf{Platforms} & \textbf{Date}  & \textbf{No. of Blobs} & \textbf{No. of Files} & \textbf{No. of Folders} & \textbf{No. of Commits} & \textbf{No. of Authors} & \textbf{No. of Projects} & \textbf{No. of Releases}\\
\midrule
Software Heritage & July 11, 2020 & NA          & 8.5B        & 7.3B          & 1.8B          & 36.4M         & 132.0M         & 14.8M          \\
GHTorrent         & June, 2019  & NA            & NA            & NA               &  1.4B             & 32.4M           &   125.4M             & NA           \\
BOA (GitHub set)  & October, 2019 & NA          & 484.9M      & NA            & 23.2M         & NA            & 7.8M           & NA             \\
WoC               & Mar, 2020     & 7.9B        & 12.4B            & 8.3B          & 2.0B          & 42.1M         & 123.8M         & NA            \\
GH Archive   & Dec, 2019 & NA & NA & NA & NA & 26.9M & 126.6M & NA\\
\bottomrule
\end{tabular}
}
\end{table}

\subsection{GHTorrent}
GHTorrent~\cite{gousios2012ghtorrent} captures GitHub's event stream
and uses the captured data to reconstruct GitHub's object model of
software interaction history that is visible to GitHub users. Its
focus is thus more on software process than content: it captures the
identities of projects, users, and commits, as well as their
interactions in the form of issues, pull requests, comments, and
their starring, labeling, and joining of projects. GHTorrent does
not capture the actual contents of files in the projects.

Its main advantage over World of Code is that it captures a
significant dimension of non-coding work that happens in projects:
bug reporting and design discussions that happen in pull request and
issue threads.  It is less suited to research involving the actual
content of code files, since it does not contain or index the files
contained in the projects. It also is limited strictly to GitHub
data.

Further, GHTorrent's method for collecting data has inherent
limitations~\cite{Kalliamvakou2014}: GitHub's event stream has
changed in format over time, has produced buggy or inconsistent data
in the past, and has had outages; thus GHTorrent provides only a
best-effort reconstruction of Github-based software development.

\subsection{Software Heritage}
Software Heritage(SH)~\cite{di2017software,msr-2019-swh} has aims and methods that are similar to that of World of Code: they aim to collect ``all publicly available source code, with all its development history,
organizing it into a canonical structure, and providing unique, intrinsic identifiers for all its parts, enabling its better fruition while
ensuring its long term preservation.''~\cite{di2017software}.  

Like WoC, this platform also collects software on an ongoing basis from multiple sources, including Github, Gitlab, the Debian package repository, Gitorious, Google code, and PyPi~\cite{msr-2019-swh}.  They collect commits, authors, repositories, directories, filenames, and the blobs themselves. Unlike WoC, they also collect branches and releases, in a cross-VC representation they call ``snapshots''.

SH holds this archive of data on a set of servers, with a web
interface designed around its role as a preservation archive -- it
is convenient for users to register new code bases and examine
historical ones one at a time, but it is not designed to allow the
public to do arbitrary research queries.  For research purposes,
they offer a large (~1TB) graph database of
metadata~\cite{msr-2019-swh}; this makes a broader range of metadata
queries possible, but the files/blobs themselves are still only
accessible one at a time, via an API; there would be no efficient
way to run analysis on the content of large numbers of blobs in
order to, for example, search for a particular method name or
language feature use.

\subsection{BOA}\label{ss:BOA}
The Boa system provides a commendable option in regards to Git
analysis. However, it does have some issues when analyzing complex
research questions due to inherent complexities related to
querying. More specifically, Boa's data storage
format~\cite{hung2020boa} requires the use of MapReduce, a complex
model used for batch processing large datasets, which becomes worse
when the desired data requires multiple uses of MapReduce.

Hung~\cite{hung2020boa} provides an interesting attempt to subvert
this issue by providing what they characterize as "Materialized
Views", which allows users to access the results of previous queries
from other users as a way to simplify the data retrieval
process. This approach helps future users for accessing data that
have been previously requested by others, however, chaining
successive MapReduces for complex tasks is still acquired for the
initial request. The WoC platform, on the hand, provides
basemappings as direct connections between various data
categories. These basemappings are stored in a constant time
database which ensures the data retrieval process maintains a
linearity in speed.

Another drawback of Boa is that it only provides datasets that were
compiled using Java and Python. Although Boa has code that supports
additional languages (e.g. PHP, XML, CSS), no datasets are provided
with these languages so far. Furthermore, Boa only provides data
associated with GitHub, and additional efforts are required if users
are want to create new datasets from a manually selected set of Git
repositories out of GitHub. The latest Boa dataset was compiled in
October 2019, and contains around 7.8M projects, 380K Code
Repositories, 23.2M Revisions, 146.4M Unique files, 485M File
snapshots, and 71.8B AST Nodes.


We implemented a speed test on the same machine (to maintain
comparability) to perform a comprehensive comparison between these
two systems. We installed the Python boa-api client on this machine,
which also has access to WoC basemappings. We took as an example
counting the total number of committers in each project, which is
explicitly described in the list of examples on the Boa official
webpage\footnote{\url{http://boa.cs.iastate.edu/examples/index.php\#scheme-use}}. Given
the substantial difference in the number of projects contained
within each platform, we chose to randomly select 7,830,023 (the
number of projects in Boa) projects from WoC for comparison. We
leveraged Perl API and project-to-author map and completed this task
at a time cost of 259 seconds. We directly ran the source code of
the example provided on Boa webpage, and finished the same task in
383 seconds. Based on our test, we believe that WoC platform can
provide better performance on specific tasks than Boa given that the
WoC selection task was run on an order-of-magnitude larger dataset.

\subsection{GH Archive}
GH Archive is a project to record the public GitHub timeline of
events, archive it, and make it easily accessible for further
analysis\footnote{\url{https://www.gharchive.org/}}. New public
GitHub events/activities are collected every hour, and have been
archived since Feb 12, 2011. Users can access these data sets either
by directly downloading each hourly compressed collection, or by
running BigQuery in Google Cloud Platform where the full data set is
stored as SQL tables.

By selecting unique repos over the nine databases representing
years from 2011 until 2019, we found that GHArchive contains around
126.6 million projects. 

GH Archive data has supported several research projects,
visualizations and talks, such as GitHub
Analytics\footnote{\url{https://github.com/harishvc/githubanalytics}}
and GitHut\footnote{\url{https://githut.info/}}. Compared to other
platforms, GH Archive has advantages for research that studies a
specific time range, requires publicly available services (cloud
service) and most recent data access (quick hourly update). However,
there are drawbacks as well: this archive contains only the log of
visible events, not the blobs and files they represent.  Parsing
these logs is nontrivial, as log formats are complex and have
changed subtly over the years; and furthermore, relationships among
development components (e.g., what are the commits made by a
specific developer) are not directly represented, and must instead
be collected and indexed by the user.

\section{ \texorpdfstring{\hl{Hackathon Event}}{Hackathon Event}}\label{s:hackathon}
In early November, 2019, we hosted a hackathon event at the Carnegie
Mellon University campus, during which a number of researchers in
software domain from all over the world gathered and participated
to further explore the potentials of WoC platform, propose missing
services that would be beneficial for their researches, and,
overall, to familiarize themselves with WoC.  In this section, we
report the hackathon event workflow, a brief summary of projects
proposed and implemented during this hackathon, and a published
research work (extension of one of the projects).

\subsection{Event Workflow}\label{ss:setup}

We hosted a pre-hackathon event virtually in October 2019, where
we shared the instructions on the setting up of WoC environment, and
went through a live tutorial of WoC together with participants. By
offering this training opportunity, we hope that the participants
could become familiar to WoC services and dataset before attending
the official event.

The official event started on the evening of Nov 1st. After a short
introduction to WoC, participants were asked to brain storm for
ideas where WoC could be leveraged to assist. These ideas were
further classified into a few categories (software risk, ecosystem
study, API usage, developer migration, infrastructure improvement,
etc.) and participants were asked to pick one of the topic at will
to contribute. In the end, four groups emerged, and each group was
expected to finalize the project road map, collaborate efficiently
and present their work by the noon of Nov 3rd.

During this event, we received a number of requests for certain
functionalities that were missing. Due to the practical impacts,
these services become our priority focus to be implemented since
then.

\subsection{Event Projects}\label{ss:summary-projects}
We briefly describes each project in the hackathon event as follows.
\begin{itemize}
    \item Bot detection~\footnote{https://github.com/ssc-oscar/BIMAN\_bot\_detection}: This group was tasked with detecting code-commits bots that are active various social-coding platforms.
    \item Bridge\footnote{\url{https://github.com/woc-hack/thebridge}}: This group consists of members from the Software Heritage project, members familiar with the GHTorrent project, and WoC. The main objective was to compare datasets across different platforms, and build a joint service/entrance as a bridge to facilitate and enable the access to these platforms.
    \item Workers Comprehension\footnote{\url{https://github.com/woc-hack/Workers-Comprehension}}: This group was tasked with determining knowledge transfer between developers when changes were made to files within a project. If multiple authors had made changes to a file it was hypothesized that the knowledge transfer between those developers could be determined by analyzing the commit messages and lines of code modified.
    \item TAP\footnote{\url{https://github.com/woc-hack/TAP}}: This project was focused on the understanding of developers trajectory on language usage, and gender distribution among languages.
\end{itemize}

\subsection{Bot Detection}\label{ss:bot}
This project was conceived during the hackathon event with the goal
of detecting bots that commit code in various social-coding
platforms. The presence of bots in datasets used to explore
questions related to empirical software engineering can be a
nuisance, since it can significantly skew the measures of
productivity, team size, project activity etc. However, detecting
bots is not an easy task, since every developer in the entire
open-source community has to be examined to determine whether that
developer is actually a bot. Due to the size of the data, the only
practical method for dealing with this problem was using a
combination of some heuristics and some fine tuning. We used a
combination of three heuristics, the result of which was combined
using an ensemble model. We call our proposed bot detection method
\textbf{BIMAN} --- \textit{Bot Identification by commit Message,
  commit Association, and author Name}.

The first heuristic was based on the observation that the bots
sometimes have the word ``bot'' in their name or email. The data
obtained from WoC had the names and email addresses of all the
developers who contributed code to OSS, so we used regular
expressions to search for the presence of the word ``bot'', and
found that sometimes, the word was there in the name, and sometimes
it was in the email ID of the authors.  Datasets like GHTorrent, GH
Archive etc. typically do not have the name and email address of the
developers, only their GitHub ID, which would likely mean that we
would have missed a number of true bots. Moreover, GHTorrent does
not have any information about the GitHub Apps, which are
responsible for making a large number of commits, and are almost
always bots. As such, having access to both the names and email
addresses of the developers, and having access to the data about the
Apps through WoC helped us broaden our search and find more bots
using this heuristic.

The second heuristic is based on the observation that bots typically
use some type of a template for creating the commit messages for
their commits. Through WoC, it is easy to get access to all commit
messages created by a developer using the map from authors to
commits (\textit{a2c}) and commits to commit contents. While this
data is available and can be obtained from GHTorrent, it is only
limited to data from GitHub, while WoC has access to data for all
OSS projects that use git, which significantly broadens the scope
for the research. Having access to WoC helped us examine the entire
OSS ecosystem (that use git) with ease and discover a number of bots
using this heuristic.

The basis of the third heuristic was the empirical observation and
hypothesis that the files modified by a commit and the projects it
is associated with will have a different pattern for the commits
created by bots and for those created by humans. To validate this
hypothesis, we collected data on a number of measures for each
developer, viz. the total number of files changed by the developer,
the number of unique file extensions modified by them, standard
deviation of number of files per commit, mean number of files per
commit, the total number of unique projects commits have been
associated with, and the median number of projects the commits have
been associated with (includes duplicates). While this data can be
easily calculated by WoC using the authors to commits
(\textit{a2c}), commits to files (\textit{c2f}), and commits to
projects (\textit{c2p}) maps, these measures would be would be
difficult to calculate quickly (or at all) for the other existing
databases.

Having access to WoC gave us the unique opportunity to apply these
heuristics on the entire OSS ecosystem in a timely manner, and
discover a number of code-commit bots. We examined more than 34
million developers who have committed code to a GitHub repository,
along with detailed information for approximately 1.6 billion
commits made by them. The final ensemble model combining these three
heuristics achieved and AUC of 0.90, and the result was published in
the Mining Software Repositories conference,
2020~\cite{dey2020detecting}. We also compiled a \textbf{dataset}
with information about 461 bots, detected by \textbf{BIMAN} and
manually verified as bots, each with more than 1,000 commits, along
with detailed information about 13,762,430 commits made by these
bots, which is available at
\datadoiBOT~\cite{tapajit_dey_2020_3694401}. An extension of the work was published in~\cite{dey2020exploratory}, and a dataset containing a mapping between bot commits, projects, files, and blobs was made available at~\cite{dey2020mapping}.

\section{Future work}\label{s:future}

To have an impact on research practice, the WoC prototype needs to be exposed via reliable services that help with research and do not overwhelm the platform. Currently, we only have Python and Perl API available. However, more languages will be supported in the future. Comparatively small pre-extracted relations will be stored into relational database to extend our accessibility to users who are used to SQL.
WoC should also accommodate additional data and computational procedures needed for discovering, correcting, cleaning, augmenting, and modeling the underlying data.
Processing hundreds of terabytes of data on powerful clusters may be out of reach for most research groups. 
Therefore, to accommodate massive queries WoC would
require more powerful hardware. Such hardware 
can be obtained from cloud vendors, but the costs of hosting and analyzing data on these platforms 
might be high. An alternative might be a few high-throughput services
that work on the hardware we currently employ. 

The differentiating features of WoC are the completeness of the collection and access to 
global relationships. Specifically, two basic services would 
be difficult to replicate outside WoC, yet be capable of high
throughput on the limited hardware. First, a reporting service that
considers prevalence of certain features, such as languages, tools,
and other technologies as well as the information about contributors
might provide services akin to those provided by a population
census. The second basic service would focus on identifying all
entities linked to a specific entity, such as files modified by a
developer, all repositories containing a specific code, or all files
that use a specific module or technology. These two capabilities, in conjunction with 
MSR technology already in use, would provide both, population-level data and complete links within entire FLOSS ecosystem. 
It would then be up to 
researchers to retrieve additional data on individual projects based on the stratified samples from the first service 
or derived from the relationships obtained from the second service.

\section{Limitations}\label{s:limitations}

We tried to make the assumptions and rationale for specific decisions clear within each section but it is important to reiterate at least some of the limitations. Despite a large size (the collection contains over 
1.45B commits), there is no guarantee it closely approximates the entirety of public version control systems as the project discovery procedure is only an approximation. Our focus on Git (due to the simplified global representation) excludes older version control systems that have not been converted to Git yet.  We regularly identify issues with data being incomplete due to collection, cleaning, or processing and we are working on an approach to continuously validate and correct it. The particular design decisions were focused on the particular computing capabilities that were available to us at the time and could/should be revisited as the prototype evolves. The entirety of research tasks that WoC provides is not exhausted by the few examples we have investigated and certain tasks may require different solutions. We do, however, think that the micro-services approach allows for simpler addition/extension/replacement of components as needs or opportunities arise than would be possible with a more monolithic architecture.

How to reliably clean, correct, integrate, and augment the collected data so that the resulting analyses accurately reflect the modeled phenomena is a concern. 
To ensure the performance of the analytics layer certain objects are 
filtered from it. For example, some of the public repositories
are created to test the performance/capabilities of Git and contain
many millions of files/blobs in a single commit. Such commits are
excluded from the analytics layer to speed-up
the commit-to-file and commit-to-blob maps. The nature of the data
may also create performance problems. For example, the most common
blob is an empty file. Mapping such blobs to all commits that create
them or to all files does not make sense, since there are millions
of commits that have created empty files. These performance-related 
modifications may affect some arguably superficial analyses, e.g., what are 
the commits with the largest number of files? We explicitly highlight 
these modifications in the WoC code to minimize potential confusion. 

Reproducibility may pose an issue in a constantly updated database. Since Git objects are added incrementally and order in which they are stored is preserved, we can reconstruct any past version of the object store. For the analytic layer, which depends on the set of Git objects available at the time, we create versions, where each of the maps described above is tagged with a version identifying the state of Git object store. Preserving these past 
versions ensures reproducibility of the results obtained from them.

The research use cases presented do not constitute an empirical evaluation of WoC usability but, instead, focus on presenting vignettes that are effective for these scenarios. Some of these vignettes went through several iterations until the simplest and fastest implementations were obtained.

\section{Conclusions}\label{s:conclusion}

We introduce WoC: a prototype of an updatable and expandable infrastructure to support research and tools that rely on version control data from the entirety of open source projects and discuss some of the research problems that require such global reach.
We discuss how we address some of the data scale and quality challenges related to data discovery, retrieval, and storage.
We enable wide data access to collected data source by providing a tool built on top of the infrastructure, which scales well with completion to query in linear time.
Furthermore, we implement ways to make this large dataset usable for a number of research tasks by doing targeted data augmentation and 
by creating data structures derived from the raw data that permit accomplishing these research tasks quickly, despite the vastness of the underlying 
data. 
Finally, we 
validated WoC by conducting actual research tasks and by inviting researchers 
to undertake investigations of their own. 
In summary, WoC can provide support for  diverse research tasks that would be otherwise out of reach for most researchers. Its focus on global properties of all public source code 
will enable research that could not be previously done and help to address highly relevant challenges
of open source ecosystem sustainability and of risks posed by this global software supply chain. 
Transforming the WoC prototype into a widely accessible platform is, therefore, our immediate priority. 

All source codes can be found in a public repository.\footnote{ https://github.com/ssc-oscar/Analytics}

\section*{Acknowledgment}

This work was supported by the National Science Foundation NSF
Awards 1633437, 1901102, and 1925615. 

\bibliography{bib}

\begin{thebibliography}{10}
\providecommand{\url}[1]{{#1}}
\providecommand{\urlprefix}{URL }
\expandafter\ifx\csname urlstyle\endcsname\relax
  \providecommand{\doi}[1]{DOI~\discretionary{}{}{}#1}\else
  \providecommand{\doi}{DOI~\discretionary{}{}{}\begingroup
  \urlstyle{rm}\Url}\fi

\bibitem{Nexus}
Nexus repository.
\newblock \url{https://www.sonatype.com/nexus-repository-oss}.
\newblock Accessed: 2019-01-02

\bibitem{abadi2009data}
Abadi, D.J.: Data management in the cloud: Limitations and opportunities.
\newblock IEEE Data Eng. Bull. \textbf{32}(1), 3--12 (2009)

\bibitem{agrawal2004integrating}
Agrawal, S., Narasayya, V., Yang, B.: Integrating vertical and horizontal
  partitioning into automated physical database design.
\newblock In: Proceedings of the 2004 ACM SIGMOD international conference on
  Management of data, pp. 359--370. ACM (2004)

\bibitem{amreen2019methodology}
Amreen, S., Bichescu, B., Bradley, R., Dey, T., Ma, Y., Mockus, A., Mousavi,
  S., Zaretzki, R.: A methodology for measuring floss ecosystems.
\newblock In: Towards Engineering Free/Libre Open Source Software (FLOSS)
  Ecosystems for Impact and Sustainability, pp. 1--29. Springer, Singapore
  (2019)

\bibitem{alfaa}
Amreen, S., Mockus, A., Bogart, C., Zhang, Y., Zaretzki, R.: Alfaa: Active
  learning fingerprint based anti-aliasing for correcting developer identity
  errors in version control data.
\newblock arXiv preprint arXiv:1901.03363  (2019)

\bibitem{amreen2020alfaa}
Amreen, S., Mockus, A., Zaretzki, R., Bogart, C., Zhang, Y.: Alfaa: Active
  learning fingerprint based anti-aliasing for correcting developer identity
  errors in version control systems.
\newblock Empirical Software Engineering pp. 1--32 (2020)

\bibitem{bajracharya2014sourcerer}
Bajracharya, S., Ossher, J., Lopes, C.: Sourcerer: An infrastructure for
  large-scale collection and analysis of open-source code.
\newblock Science of Computer Programming \textbf{79}, 241--259 (2014)

\bibitem{bevan2005facilitating}
Bevan, J., Whitehead~Jr, E.J., Kim, S., Godfrey, M.: Facilitating software
  evolution research with kenyon.
\newblock ACM SIGSOFT software engineering notes \textbf{30}(5), 177--186
  (2005)

\bibitem{Bird:2006:MES:1137983.1138016}
Bird, C., Gourley, A., Devanbu, P., Gertz, M., Swaminathan, A.: Mining email
  social networks.
\newblock In: Proceedings of the 2006 International Workshop on Mining Software
  Repositories, MSR '06, pp. 137--143. ACM, New York, NY, USA (2006).
\newblock \doi{10.1145/1137983.1138016}.
\newblock \urlprefix\url{http://doi.acm.org/10.1145/1137983.1138016}

\bibitem{bird2009promises}
Bird, C., Rigby, P.C., Barr, E.T., Hamilton, D.J., German, D.M., Devanbu, P.:
  The promises and perils of mining git.
\newblock In: Mining Software Repositories, 2009. MSR'09. 6th IEEE
  International Working Conference on, pp. 1--10. IEEE (2009)

\bibitem{budde1992prototyping}
Budde, R., Kautz, K., Kuhlenkamp, K., Z{\"u}llighoven, H.: Prototyping.
\newblock In: Prototyping, pp. 33--46. Springer (1992)

\bibitem{chacon2014pro}
Chacon, S., Straub, B.: Pro git.
\newblock Springer Nature (2014)

\bibitem{Coelho2018}
Coelho, J., Valente, M.T., Silva, L.L., Shihab, E.: Identifying unmaintained
  projects in github.
\newblock In: Proceedings of the 12th ACM/IEEE International Symposium on
  Empirical Software Engineering and Measurement. ACM (2018)

\bibitem{czerwonka2013codemine}
Czerwonka, J., Nagappan, N., Schulte, W., Murphy, B.: Codemine: Building a
  software development data analytics platform at microsoft.
\newblock IEEE software \textbf{30}(4), 64--71 (2013)

\bibitem{dey2020representation}
Dey, T., Karnauch, A., Mockus, A.: Representation of developer expertise in
  open source software.
\newblock arXiv preprint arXiv:2005.10176  (2020)

\bibitem{dey2019patterns}
Dey, T., Ma, Y., Mockus, A.: Patterns of effort contribution and demand and
  user classification based on participation patterns in npm ecosystem.
\newblock arXiv preprint arXiv:1907.06538  (2019)

\bibitem{dey2018software}
Dey, T., Mockus, A.: Are software dependency supply chain metrics useful in
  predicting change of popularity of npm packages?
\newblock In: Proceedings of the 14th International Conference on Predictive
  Models and Data Analytics in Software Engineering, pp. 66--69. ACM (2018)

\bibitem{dey2018modeling}
Dey, T., Mockus, A.: Modeling relationship between post-release faults and
  usage in mobile software.
\newblock In: Proceedings of the 14th International Conference on Predictive
  Models and Data Analytics in Software Engineering, pp. 56--65 (2018)

\bibitem{dey2020dataset}
Dey, T., Mockus, A.: A dataset of pull requests and a trained random forest
  model for predicting pull request acceptance (2020).
\newblock \doi{10.5281/zenodo.3858046}.
\newblock \urlprefix\url{https://doi.org/10.5281/zenodo.3858046}

\bibitem{dey2020deriving}
Dey, T., Mockus, A.: Deriving a usage-independent software quality metric.
\newblock Empirical Software Engineering \textbf{25}(2), 1596--1641 (2020)

\bibitem{dey2020effect}
Dey, T., Mockus, A.: Effect of technical and social factors on pull request
  quality for the npm ecosystem.
\newblock In: Proceedings of the 14th ACM/IEEE International Symposium on
  Empirical Software Engineering and Measurement (ESEM), pp. 1--11 (2020)

\bibitem{dey2020pull}
Dey, T., Mockus, A.: Which pull requests get accepted and why? a study of
  popular npm packages.
\newblock arXiv preprint arXiv:2003.01153  (2020)

\bibitem{tapajit_dey_2020_3694401}
Dey, T., Mousavi, S., Ponce, E., Fry, T., Vasilescu, B., Filippova, A., Mockus,
  A.: A dataset of bot commits (2020).
\newblock \doi{10.5281/zenodo.3610205}.
\newblock \urlprefix\url{https://doi.org/10.5281/zenodo.3610205}

\bibitem{dey2020detecting}
Dey, T., Mousavi, S., Ponce, E., Fry, T., Vasilescu, B., Filippova, A., Mockus,
  A.: Detecting and characterizing bots that commit code.
\newblock In: Proceedings of the 17th International Conference on Mining
  Software Repositories, MSR '20, p. 209–219. Association for Computing
  Machinery, New York, NY, USA (2020).
\newblock \doi{10.1145/3379597.3387478}.
\newblock \urlprefix\url{https://doi.org/10.1145/3379597.3387478}

\bibitem{dey2020mapping}
Dey, T., Vasilescu, B., Mockus, A.: {A mapping between Bot Commit, Projects,
  Files, and Blobs} (2020).
\newblock \doi{10.5281/zenodo.3699665}.
\newblock \urlprefix\url{https://doi.org/10.5281/zenodo.3699665}

\bibitem{dey2020exploratory}
Dey, T., Vasilescu, B., Mockus, A.: An exploratory study of bot commits.
\newblock arXiv preprint arXiv:2003.07961  (2020)

\bibitem{di2017software}
{Di Cosmo}, R., Zacchiroli, S.: Software heritage: Why and how to preserve
  software source code.
\newblock In: iPRES 2017: 14th International Conference on Digital
  Preservation. Kyoto, Japan (2017).
\newblock
  \urlprefix\url{https://www.softwareheritage.org/wp-content/uploads/2020/01/ipres-2017-swh.pdf
  https://hal.archives-ouvertes.fr/hal-01590958}

\bibitem{ducasse2005moose}
Ducasse, S., G{\^\i}rba, T., Nierstrasz, O.: Moose: an agile reengineering
  environment.
\newblock In: ACM SIGSOFT Software engineering notes, vol.~30, pp. 99--102. ACM
  (2005)

\bibitem{Dyer-13}
Dyer, R.: Task fusion: Improving utilization of multi-user clusters.
\newblock In: Proceedings of the 2013 companion publication for conference on
  Systems, programming, \& applications: software for humanity, SPLASH SRC, pp.
  117--118 (2013)

\bibitem{Dyer-Nguyen-Rajan-Nguyen-13}
Dyer, R., Nguyen, H.A., Rajan, H., Nguyen, T.N.: Boa: A language and
  infrastructure for analyzing ultra-large-scale software repositories.
\newblock In: Proceedings of the 35th International Conference on Software
  Engineering, {ICSE}'13, pp. 422--431 (2013)

\bibitem{Dyer-Nguyen-Rajan-Nguyen-asd}
Dyer, R., Nguyen, H.A., Rajan, H., Nguyen, T.N.: Boa: an enabling language and
  infrastructure for ultra-large scale msr studies.
\newblock The Art and Science of Analyzing Software Data pp. 593--621 (2015)

\bibitem{Dyer-Nguyen-Rajan-Nguyen-15}
Dyer, R., Nguyen, H.A., Rajan, H., Nguyen, T.N.: Boa: Ultra-large-scale
  software repository and source-code mining.
\newblock ACM Trans. Softw. Eng. Methodol. \textbf{25}(1), 7:1--7:34 (2015)

\bibitem{Dyer-Rajan-Nguyen-13}
Dyer, R., Rajan, H., Nguyen, T.N.: Declarative visitors to ease fine-grained
  source code mining with full history on billions of {AST} nodes.
\newblock In: Proceedings of the 12th International Conference on Generative
  Programming: Concepts \& Experiences, GPCE, pp. 23--32 (2013)

\bibitem{eastlake2001us}
Eastlake~3rd, D., Jones, P.: Us secure hash algorithm 1 (sha1).
\newblock Tech. rep. (2001)

\bibitem{fry2020dataset}
Fry, T., Dey, T., Karnauch, A., Mockus, A.: A dataset and an approach for
  identity resolution of 38 million author ids extracted from 2b git commits.
\newblock arXiv preprint arXiv:2003.08349  (2020)

\bibitem{german2003automating}
German, D., Mockus, A.: Automating the measurement of open source projects.
\newblock In: Proceedings of the 3rd workshop on open source software
  engineering, pp. 63--67. University College Cork Cork Ireland (2003)

\bibitem{gorton2016software}
Gorton, I., Bener, A.B., Mockus, A.: Software engineering for big data systems.
\newblock IEEE Software \textbf{33}(2), 32--35 (2016)

\bibitem{Gousi13}
Gousios, G.: The ghtorrent dataset and tool suite.
\newblock In: Proceedings of the 10th Working Conference on Mining Software
  Repositories, MSR '13, pp. 233--236. IEEE Press, Piscataway, NJ, USA (2013).
\newblock \urlprefix\url{http://dl.acm.org/citation.cfm?id=2487085.2487132}

\bibitem{gousios2014exploratory}
Gousios, G., Pinzger, M., Deursen, A.v.: An exploratory study of the pull-based
  software development model.
\newblock In: Proceedings of the 36th International Conference on Software
  Engineering, pp. 345--355. ACM (2014)

\bibitem{gousios2009alitheia}
Gousios, G., Spinellis, D.: Alitheia core: An extensible software quality
  monitoring platform.
\newblock In: Software Engineering, 2009. ICSE 2009. IEEE 31st International
  Conference on, pp. 579--582. IEEE (2009)

\bibitem{gousios2012ghtorrent}
Gousios, G., Spinellis, D.: Ghtorrent: Github's data from a firehose.
\newblock In: Mining software repositories (msr), 2012 9th ieee working
  conference on, pp. 12--21. IEEE (2012)

\bibitem{gousios2014lean}
Gousios, G., Vasilescu, B., Serebrenik, A., Zaidman, A.: Lean ghtorrent: Github
  data on demand.
\newblock In: Proceedings of the 11th working conference on mining software
  repositories, pp. 384--387. ACM (2014)

\bibitem{gousios2014dataset}
Gousios, G., Zaidman, A.: A dataset for pull-based development research.
\newblock In: Proceedings of the 11th Working Conference on Mining Software
  Repositories, pp. 368--371. ACM (2014)

\bibitem{HMPQ10}
Hackbarth, R., Mockus, A., Palframan, J., Weiss, D.: Assessing the state of
  software in a large enterprise.
\newblock Journal of Empirical Software Engineering \textbf{10}(3), 219--249
  (2010)

\bibitem{howison2006flossmole}
Howison, J., Conklin, M., Crowston, K.: Flossmole: A collaborative repository
  for floss research data and analyses.
\newblock International Journal of Information Technology and Web Engineering
  (IJITWE) \textbf{1}(3), 17--26 (2006)

\bibitem{hung2020boa}
Hung, C.S., Dyer, R.: Boa views: Easy modularization and sharing of msr
  analyses

\bibitem{Kalliamvakou2014}
Kalliamvakou, E., Gousios, G., Blincoe, K., Singer, L., German, D.M., Damian,
  D.: The promises and perils of mining github.
\newblock In: Proceedings of the 11th Working Conference on Mining Software
  Repositories, MSR 2014, p. 92–101. Association for Computing Machinery, New
  York, NY, USA (2014).
\newblock \doi{10.1145/2597073.2597074}.
\newblock \urlprefix\url{https://doi.org/10.1145/2597073.2597074}

\bibitem{TaRe}
Kim, S., Zimmermann, T., Kim, M., Hassan, A.E., Mockus, A., G\^{i}rba, T.,
  Pinzger, M., Jr., E.J.W., Zeller, A.: Ta-re: an exchange language for mining
  software repositories.
\newblock In: ICSE'06 Workshop on Mining Software Repositories, pp. 22--25.
  Shanghai, China (2006).
\newblock \urlprefix\url{http://dl.acm.org/authorize?804411}

\bibitem{doc2vec}
Le, Q., Mikolov, T.: Distributed representation of sentences and documents.
\newblock In: Proceedings of the 31 st International Conference on Machine
  Learning, vol.~32. JMLR, Beijing,China (2014).
\newblock \urlprefix\url{https://cs.stanford.edu/~quocle/paragraph\_vector.pdf}

\bibitem{leavitt2010will}
Leavitt, N.: Will nosql databases live up to their promise?
\newblock Computer \textbf{43}(2) (2010)

\bibitem{lichter1994prototyping}
Lichter, H., Schneider-Hufschmidt, M., Zullighoven, H.: Prototyping in
  industrial software projects-bridging the gap between theory and practice.
\newblock IEEE transactions on software engineering \textbf{20}(11), 825--832
  (1994)

\bibitem{ma2016crowdsourcing}
Ma, Y., Dey, T., Smith, J.M., Wilder, N., Mockus, A.: Crowdsourcing the
  discovery of software repositories in an educational environment.
\newblock PeerJ Preprints \textbf{4}, e2551v1

\bibitem{ma2019methodology}
Ma, Y., Mockus, A., Zaretzki, R., Bradley, R., Bichescu, B.: A methodology for
  analyzing uptake of software technologies among developers.
\newblock arXiv preprint arXiv:1908.11431  (2019)

\bibitem{M07}
Mockus, A.: Large-scale code reuse in open source software.
\newblock In: ICSE'07 Intl. Workshop on Emerging Trends in FLOSS Research and
  Development. Minneapolis, Minnesota (2007).
\newblock \urlprefix\url{papers/ossreuse.pdf}

\bibitem{M09msr}
Mockus, A.: Amassing and indexing a large sample of version control systems:
  towards the census of public source code history.
\newblock In: 6th IEEE Working Conference on Mining Software Repositories
  (2009).
\newblock \urlprefix\url{papers/amassing.pdf}

\bibitem{M14}
Mockus, A.: Engineering big data solutions.
\newblock In: ICSE'14 FOSE (2014).
\newblock \urlprefix\url{papers/BigData.pdf}

\bibitem{moniruzzaman2013nosql}
Moniruzzaman, A., Hossain, S.A.: Nosql database: New era of databases for big
  data analytics-classification, characteristics and comparison.
\newblock arXiv preprint arXiv:1307.0191  (2013)

\bibitem{munaiah2017curating}
Munaiah, N., Kroh, S., Cabrey, C., Nagappan, M.: Curating github for engineered
  software projects.
\newblock Empirical Software Engineering \textbf{22}(6), 3219--3253 (2017)

\bibitem{ossher2009sourcererdb}
Ossher, J., Bajracharya, S., Linstead, E., Baldi, P., Lopes, C.: Sourcererdb:
  An aggregated repository of statically analyzed and cross-linked open source
  java projects.
\newblock In: Mining Software Repositories, 2009. MSR'09. 6th IEEE
  International Working Conference on, pp. 183--186. IEEE (2009)

\bibitem{msr-2019-swh}
Pietri, A., Spinellis, D., Zacchiroli, S.: The software heritage graph dataset:
  Public software development under one roof.
\newblock In: Proceedings of the 16th International Conference on Mining
  Software Repositories, MSR '19, pp. 138--142. IEEE Press (2019).
\newblock \doi{10.1109/MSR.2019.00030}.
\newblock
  \urlprefix\url{https://www.softwareheritage.org/wp-content/uploads/2020/01/msr-2019-swh.pdf
  https://upsilon.cc/~zack/research/publications/msr-2019-swh.pdf}

\bibitem{qi2007fast}
Qi, Z.: Fast sha1 implementation (2007).
\newblock US Patent 7,299,355

\bibitem{boa-website}
Rajan, H., Nguyen, T.N., Dyer, R., Nguyen, H.A.: Boa website.
\newblock http://boa.cs.iastate.edu/ (2015)

\bibitem{rosch1999principles}
Rosch, E.: Principles of categorization.
\newblock Concepts: core readings \textbf{189} (1999)

\bibitem{rozenberg2016comparing}
Rozenberg, D., Beschastnikh, I., Kosmale, F., Poser, V., Becker, H., Palyart,
  M., Murphy, G.C.: Comparing repositories visually with repograms.
\newblock In: Proceedings of the 13th International Conference on Mining
  Software Repositories, pp. 109--120. ACM (2016)

\bibitem{russom2011big}
Russom, P., et~al.: Big data analytics.
\newblock TDWI best practices report, fourth quarter \textbf{19}(4), 1--34
  (2011)

\bibitem{Promise}
Sayyad~Shirabad, J., Menzies, T.: {The {PROMISE} Repository of Software
  Engineering Databases.}
\newblock School of Information Technology and Engineering, University of
  Ottawa, Canada (2005).
\newblock \urlprefix\url{http://promise.site.uottawa.ca/SERepository}

\bibitem{spinellis2020dataset}
Spinellis, D., Kotti, Z., Mockus, A.: A dataset for github repository
  deduplication.
\newblock arXiv preprint arXiv:2002.02314  (2020)

\bibitem{tiwari2017candoia}
Tiwari, N.M., Upadhyaya, G., Nguyen, H.A., Rajan, H.: Candoia: A platform for
  building and sharing mining software repositories tools as apps.
\newblock In: MSR'17: 14th International Conference on Mining Software
  Repositories (2017)

\bibitem{tiwari2016candoia}
Tiwari, N.M., Upadhyaya, G., Rajan, H.: Candoia: A platform and ecosystem for
  mining software repositories tools.
\newblock In: Proceedings of the 38th International Conference on Software
  Engineering Companion, pp. 759--764. ACM (2016)

\bibitem{upadhyaya2017accelerating}
Upadhyaya, G., Rajan, H.: On accelerating ultra-large-scale mining.
\newblock In: Proceedings of the 39th International Conference on Software
  Engineering: New Ideas and Emerging Results Track, pp. 39--42. IEEE Press
  (2017)

\bibitem{upadhyaya2018accelerating}
Upadhyaya, G., Rajan, H.: On accelerating source code analysis at massive
  scale.
\newblock IEEE Transactions on Software Engineering  (2018)

\bibitem{wang2005collision}
Wang, X., Yin, Y.L., Yu, H.: Collision search attacks on sha1 (2005)

\bibitem{Winkler}
Winkler, W.: String comparator metrics and enhanced decision rules in the
  fellegi-sunter model of record linkage  (1990)

\bibitem{Winkler06overviewof}
Winkler, W.E.: Overview of record linkage and current research directions.
\newblock Tech. rep., BUREAU OF THE CENSUS (2006)

\end{thebibliography}
\bibliographystyle{spmpsci}

\begin{appendices}
\section{Source Code for Cmt2ATShow.perl}\label{apx:source code}
\begin{lstlisting}[language=perl]
#!/usr/bin/perl -I /home/audris/lib64/perl5 -I /home/audris/lib/x86_64-linux-gnu/perl
use strict;
use warnings;
use Error qw(:try);
use TokyoCabinet;
use Compress::LZF;

sub toHex { 
  return unpack "H*", $_[0]; 
} 
sub fromHex { 
  return pack "H*", $_[0]; 
}

my $split = 1;
$split = $ARGV[1] + 0 if defined $ARGV[1];

my %c2at;
for my $sec (0..($split-1)){
  my $fname = "$ARGV[0].$sec.tch";
  $fname = $ARGV[0] if ($split == 1);
  tie %{$c2at{$sec}}, "TokyoCabinet::HDB", "$fname", TokyoCabinet::HDB::OREADER,   
         16777213, -1, -1, TokyoCabinet::TDB::TLARGE, 100000
      or die "cant open $fname\n";
}

while (<STDIN>){
  chop ();
  my $c = fromHex($_);
  my $ss = pack 'H*', substr ($_, 0, 2);
  my $sec = (unpack "C", $ss)%$split;
  if (defined $c2at{$sec}{$c}) {
    my ($time, $author) = split(/;/, $c2at{$sec}{$c});
    my @parts = localtime($time);
    my $year= $parts[5] + 1900;
    print $year.";".$author."\n";
  }
}
for my $sec (0..($split-1)){
  untie %{$c2at{$sec}};
}
\end{lstlisting}

\section{Source Code for the custom \texttt{lsort} command in tutorial}\label{lsort}
\begin{lstlisting}[language=bash]
#!/bin/bash
export LC_ALL=C 
export LANG=C  
sz=${1:-10G}
shift
sort -T. -S $sz --compress-program=gzip $@
\end{lstlisting}{}
\end{appendices}

\end{document}